\documentclass{aa}
\makeatletter
\renewcommand*\aa@pageof{, page \thepage{} of \pageref*{LastPage}}
\makeatother
\usepackage[titletoc,title]{appendix}
\usepackage{graphicx}
\usepackage{float}
\usepackage{placeins}
\usepackage{txfonts}
\usepackage{hyperref}
\usepackage{color}
\usepackage{xcolor}

\begin{document}

\title{Measuring galaxy cluster mass profiles into the low acceleration regime with galaxy kinematics}
\titlerunning{Cluster dynamics from galaxy kinematics}
\authorrunning{P. Li et al.}

   \author{Pengfei Li
          \inst{1}\fnmsep\thanks{Humboldt fellow.}
          \and
          Yong Tian\inst{2}
          \and
          Mariana P. Júlio\inst{1,3}
          \and
          Marcel S. Pawlowski\inst{1}
          \and
          Federico Lelli\inst{4}
          \and
          Stacy S. McGaugh\inst{5}
          \and
          James M. Schombert\inst{6}
          \and
          Justin I. Read\inst{7}
          \and
          Po-Chieh Yu\inst{8}
          \and
          Chung-Ming Ko\inst{2}
          }

   \institute{Leibniz-Institute for Astrophysics,
              An der Sternwarte 16, 14482 Potsdam, Germany\\
              \email{pli@aip.de}
              \and
              Institute of Astronomy, National Central University, Taoyuan 32001, Taiwan
              \and
              Institut f\"{u}r Physik und Astronomie, Universit\"{a}t Potsdam, Karl-Liebknecht-Straße 24/25, D-14476 Potsdam, Germany
              \and
              INAF – Arcetri Astrophysical Observatory, Largo Enrico Fermi 5, I-50125, Firenze, Italy
              \and
              Department of Astronomy, Case Western Reserve University, 10900 Euclid Avenue, Cleveland, OH 44106, USA
              \and
              Department of Physics, University of Oregon, Eugene, OR 97403, USA
              \and
              Department of Physics, University of Surrey, Guildford GU2 7XH, UK
              \and
              College of General Studies, Yuan-Ze University, Taoyuan 32003, Taiwan
             }

   \date{Received xxx; accepted xxx}
 
  \abstract
{We probe the dynamical mass profiles of 10 galaxy clusters from the HIghest X-ray FLUx Galaxy Cluster Sample (HIFLUGCS) using galaxy kinematics. We numerically solve the spherical Jeans equation, and parameterize the dynamical mass profile and the galaxy velocity anisotropy profile using two general functions to ensure that our results are not biased towards any specific model. The mass-velocity anisotropy degeneracy is ameliorated by using two `virial shape parameters’ that depend on the fourth moment of velocity distribution. The resulting velocity anisotropy estimates consistently show a nearly isotropic distribution in the inner regions, with an increasing radial anisotropy towards large radii. We compare our derived dynamical masses with those calculated from X-ray gas data assuming hydrostatic equilibrium, finding that massive and rich relaxed clusters generally present consistent mass measurements, while unrelaxed or low-richness clusters have systematically larger total mass than hydrostatic mass by an average of 50\%. This might help alleviate current tensions in the measurement of $\sigma_8$, but it also leads to cluster baryon fractions below the cosmic value. Finally, our approach probes accelerations as low as $10^{-11}$ m s$^{-2}$, comparable to the outskirts of individual late-type galaxies. We confirm that galaxy clusters deviate from the radial acceleration relation defined by galaxies.
}

   \maketitle

%________________________________________________________________

\section{Introduction}

Clusters of galaxies present the most prominent structures in the Universe given they are the largest self-gravitating objects. Their distribution and abundance accommodate critical information on the geometry of the Universe and the growth of structures \citep{Bohringer2014, Planck2016XXIV}, which often require measurements of their total mass. Robust estimates of cluster masses rely on constraints from spatially resolved dynamics, which also makes clusters excellent laboratories for testing various dark matter (DM) models and alternative theories of gravity \citep[e.g.][]{Angus2008}. 

Hot X-ray emitting gas under the assumption of hydrostatic equilibrium is the most commonly used dynamical tracer to derive the dynamical mass profiles of galaxy clusters, given that they are generally X-ray selected and their X-ray surface brightness profiles are usually available. Recent observations have managed to measure radially varying temperature profiles \citep[e.g.][]{Ghirardini2019, Liu2023}, resulting in more accurate estimates for the total mass. The assumption of hydrostatic equilibrium is sometimes in question due to the existence of non-thermal pressure \citep{Lau2009,Nelson2014}. Perhaps more interestingly, the cosmological constraints from cluster counts \citep{Planck2016XXIV} is in tension with that from the cosmic microwave background \citep[CMB,][]{Planck2016XIII}, which can be resolved by introducing a 40\% bias in mass. These discrepancies have motivated studies comparing hydrostatic mass with that measured with other approaches, usually gravitational lensing. Gravitational lensing \citep{Postman2012, Umetsu2016} measures the deflection of photons from background sources which does not assume dynamical equilibrium. As such, it is considered the most reliable method, but it has only be applied to a subset of galaxy clusters and subject to projection effects.

Historically, using cluster galaxies as dynamical tracers was the very first method to measure the dynamical mass of galaxy clusters and uncover their DM content \citep{Zwicky1933}. However, this approach is relatively less developed and applied, because it requires expensive spectroscopic measurements for a large number of galaxies. On the technical side, it is also challenging to derive 3D dynamical mass profiles from projected galaxy distributions and line-of-sight velocities. There are three mass estimators that have been used in the literature: virial theorem, caustic approach, Jeans formula equation. Employing the virial theorem \citep{Limber1960, Biviano2006} is the easiest, but it only provides the estimate of the virial mass. The caustic approach \citep{Diaferio1997, Diaferio1999, Serra2011} considers the fact that gravitationally bounded galaxies cannot exceed the escaping velocity, so one would expect a trumpet-shape distribution of galaxies in the projected phase space, given the gravitational field decreases towards large radii. This approach does not rely on dynamical equilibrium. The Jeans equation \citep{vanderMarel1994} relates the phase-space distribution of galaxies to the total gravitational potential. It assumes that galaxies are in dynamical equilibrium. Deviations from the equilibrium are typically accompanied by disturbed gas distributions and thereby can be partially identified from X-ray images \citep{Nagai2007}. Both the caustic approach and the Jeans modeling have to deal with the degeneracy between the density profile and the velocity anisotropy. Strategies that have been implemented include assuming a constant velocity anisotropy \citep[e.g.][]{Biviano2013}, or parametrizing its profile with a single parameter. The latter generally assumes perfect isotropy either in the center \citep{Mamon2005II,Tiret2007} or at the scale radius of DM halos \citep{Biviano2013}. These inevitably introduce some biases. \citet{Foex2017} compared the cluster mass estimated with these three approaches and found they are generally higher than hydrostatic mass by 20-50\%.

With more spectra of cluster galaxies becoming available and especially the advent of SDSS-V \citep{Almeida2023}, it is now worth further developing the approach with galaxy kinematics. In this paper, we numerically solve the Jeans equation and adopt more general functions to describe the dynamical mass profile and velocity anisotropy. Our results are thus not biased towards any specific halo model. We ameliorate the degeneracy between the density profile and velocity anisotropy with fourth moment of velocity distribution, i.e. two virial shape parameters. We test the approach using 16 HIFLUGCS clusters and compare with their hydrostatic mass profiles. Section 2 describes our cluster sample; Section 3 introduces the approach we employed; Section 4 presents the comparison with hydrostatic mass for six disturbed clusters; Section 5 presents the detailed dynamical analysis for ten X-ray relaxed clusters. We discuss our results and conclude the paper in Section 6.

\section{HIFLUGCS clusters}

\subsection{X-ray data}

In this paper, we adopted a subsample of HIFLUGCS clusters, which were observed by both the ROSAT All-Sky Survey and XMM-Newton, leading to high-quality X-ray data \citep{Zhang2011}. The complete sample of 63 clusters is X-ray selected and constructed by \citet{Reiprich2002}. \citet{Chen2007} provide fits of their surface brightness profiles with the $\beta$ model \citep{Cavaliere1976},
\begin{equation}
    S(R) = S_0\Big(1+R^2/r_c^2\Big)^{-3\beta+\frac{1}{2}},
    \label{eq:betaSB}
\end{equation}
where $R$ is the projected radius, $r_c$ is the core radius, $S_0$ is the central surface brightness, and $\beta$ is the power index. Assuming spherical symmetry, the above function can be de-projected to derive the electron number density profile assuming that X-rays are produced by thermal bremsstrahlung,
\begin{equation}
    n_e = n_c\Big(1+r^2/r_c^2\Big)^{-3\beta/2}
\end{equation}
where $r$ is the 3D radius, and $n_c$ is the central electron number density. \citet{Chen2007} adopted $H_0=50$ km s$^{-1}$ Mpc$^{-1}$ for distance estimates, which is significantly lower than current values. In order to maintain consistency with current estimates, we adopt $H_0 = 70$ km s$^{-1}$ Mpc$^{-1}$ and re-scale his results for core radius and electron number density: $n_c = n_{\rm c,original}/1.4$ and $r_c=1.4r_{\rm c,original}$. The enclosed gas mass profile is then
\begin{equation}
    M_{\rm gas}(<r) = 4\pi\frac{A}{Z}m_pn_cr_c^3\int_0^x(1+x^2)^{-\frac{3}{2}\beta}x^2{\rm d}x,
\end{equation}
where $x=r/r_c$; $m_p$ is the mass of protons; the mean nuclear mass and charge numbers are $A\simeq1.4$ and $Z\simeq1.2$ for the intracluster medium (ICM) with 0.3 solar abundance \citep{Anders1989}. Assuming the ICM is in hydrostatic equilibrium and isothermal, the dynamical mass profile can be derived through
\begin{equation}
    M_{\rm hydro}(<r) = \frac{3\beta kT_hr}{G\mu m_p}\frac{(r/r_c)^2}{1+(r/r_c)^2},
    \label{eq:hydromass}
\end{equation}
where $T_h$ is the temperature of the hotter bulk component in the two-phase model. The cooler phase is included to account for the possible cooling core. \citet{Chen2007} pointed out $T_h$ provides a better measure for the gravitational potential and total mass than the single emission-weighted temperature $T_m$. We, therefore, use $T_h$ when deriving the hydrostatic mass in this paper. The mean molecular weight $\mu$ is given by
\begin{equation}
    \mu=\rho_{\rm gas}/(m_pn_{\rm gas}) \simeq (2X + 0.75Y + 0.56Z)^{-1},
\end{equation}
where $n_{\rm gas}$ is the total number density including electrons, protons, ionized Helium, and other ionized elements; $X$, $Y$ and $Z$ are the mass fractions of Hydrogen, Helium, and other elements, respectively. For 0.3 solar abundance, $X\simeq0.716$ and $Y\simeq0.278$. This gives $\mu\simeq0.6$.

\subsection{Optical data}
The collection of high-quality spectro-photometric data for the HIFLUGCS clusters is an achievement resulting from decades of dedicated observations, including the compilations in \citet{Andernach2005, Zhang2011}. In this paper, we utilize the compiled sample from \citet{Tian2021}, which includes the comprehensive memberships assembled in the literature and organized from SIMBAD \citep{Wenger2000}. This sample is cleaned by excluding uncertain or repeated members. Each cluster in the sample has a single brightest cluster galaxy (BCG), and its position is defined as the optical center. \citet{Tian2021} required the offset between the optical center and X-ray weighted center to be smaller than 60 kpc. Therefore, we do not distinguish between optical or X-ray centers in this paper.

In order to effectively constrain dynamical mass profiles, we require at least three radial bins for each cluster with at least 25 galaxies in each bin. We, therefore, exclude clusters with less than 75 corfirmed member galaxies. We also remove two galaxy groups, NGC 4636 and A1060, from our cluster sample. In the end, we retain 16 clusters in the analysis. According to \citet{Zhang2011}, six of them are disturbed, so their galaxy kinematics may not trace the gravitational potential. Therefore, we will study disturbed and undisturbed clusters separately.

\section{Methodology}

\subsection{Galaxy kinematics}
Using galaxy kinematics to determine the dynamical mass of clusters assumes that galaxies obey the collisionless Boltzmann equation \citep{Binney2008},
\begin{equation}
    \frac{{\rm d}f}{{\rm d}t} = \frac{\partial f}{\partial t} + \nabla_xf\cdot\vec{v} - \nabla_vf\cdot\nabla_x\Phi=0,
\end{equation}
where $f(\vec{x},\vec{v})$ is the distribution function of galaxies, and $\Phi$ is the total gravitational potential given by $\nabla^2 \Phi = 4\pi G \rho$. The dynamical equilibrium is a critical assumption. It is also a central question we aim to answer by comparing the results with those using hydrostatic equilibrium. Since we assume spherical symmetry for all the considered clusters, the Boltzmann equation can be simplified as the spherical Jeans equation \citep{Binney2008},
\begin{equation}
    \frac{1}{\nu}\frac{\partial}{\partial r}(\nu\sigma^2_r) + \frac{2\beta(r)\sigma^2_r}{r} = -\frac{GM(<r)}{r^2},
    \label{eq:SphereJeans}
\end{equation}
where $\nu=\int f{\rm d^3}v$ is the number density of tracer galaxies, $\sigma_r$ is the radial velocity dispersion, $\beta=1-\sigma^2_t/\sigma^2_r$ is the velocity anisotropy with $\sigma_t$ the tangential anisotropy, and $M(<r)$ is the enclosed total mass. Once $\nu$, $\sigma_r$ and $\beta$ are determined, equation \ref{eq:SphereJeans} provides a direct measurement for the enclosed total mass. In fact, it is more common to use the integrated equation \citep{vanderMarel1994},
\begin{equation}
    \sigma^2_r = \frac{1}{\nu(r)g(r)}\int_r^\infty\frac{GM(<r')\nu(r')}{r'^2}g(r'){\rm d}r',
\end{equation}
where $g(r)=\exp{(\int\frac{2\beta(r)}{r}{\rm d}r)}$. In observations, only projected distributions of galaxies ($\Sigma_{\rm gal}$) and line-of-sight velocity dispersion ($\sigma_{\rm los}$) are observable. Therefore, it is important to have an equation relating $\sigma_{\rm los}$ to $\sigma_{\rm r}$, which has been derived by \citet{Binney1982},
\begin{equation}
    \sigma^2_{\rm los}(R) = \frac{2}{\Sigma_{\rm gal}(R)}\int_R^\infty\Big(1-\beta\frac{R^2}{r^2}\Big)\frac{\nu(r)\sigma^2_rr}{\sqrt{r^2-R^2}}{\rm d}r.
    \label{eq:sigmalos}
\end{equation}

\subsection{GravSphere code}

\begin{figure*}
    \centering
    \includegraphics[scale=0.45]{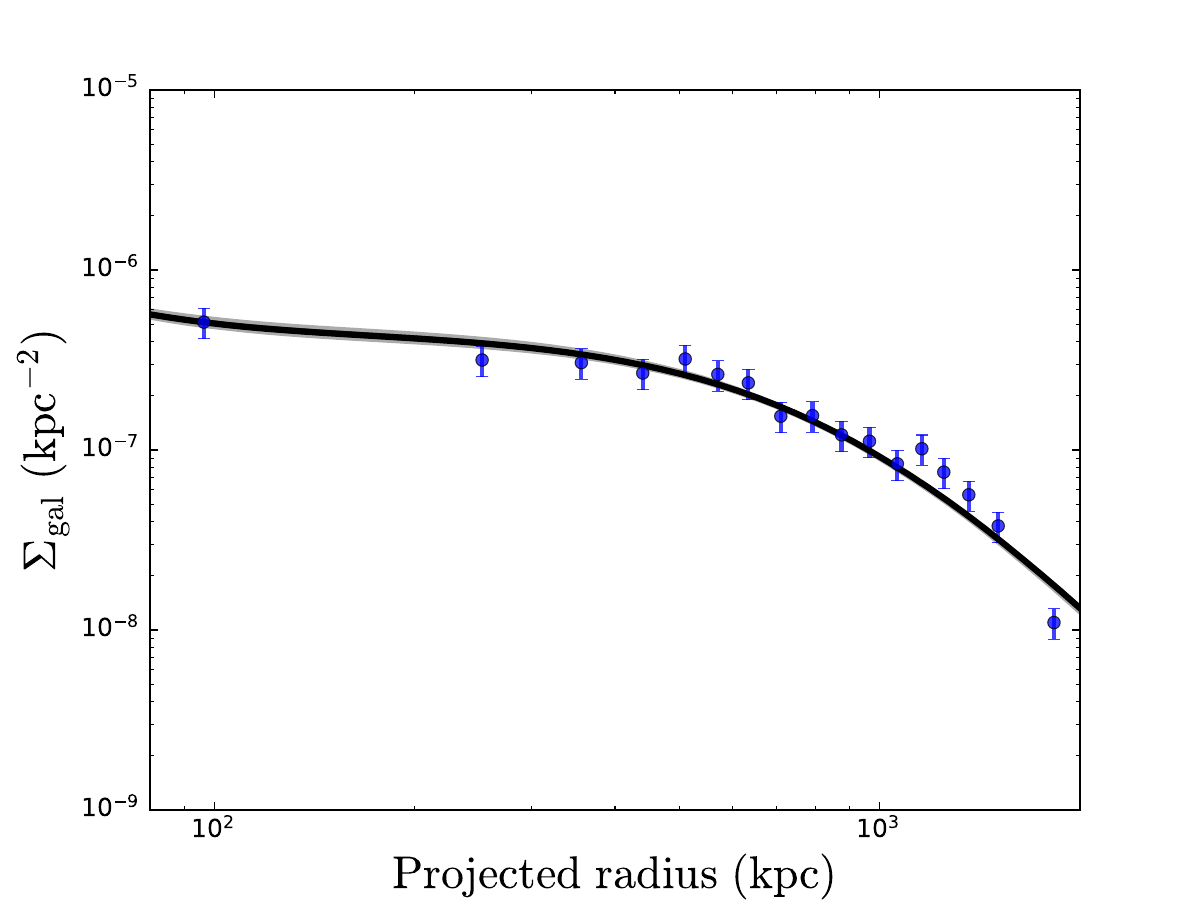}\includegraphics[scale=0.45]{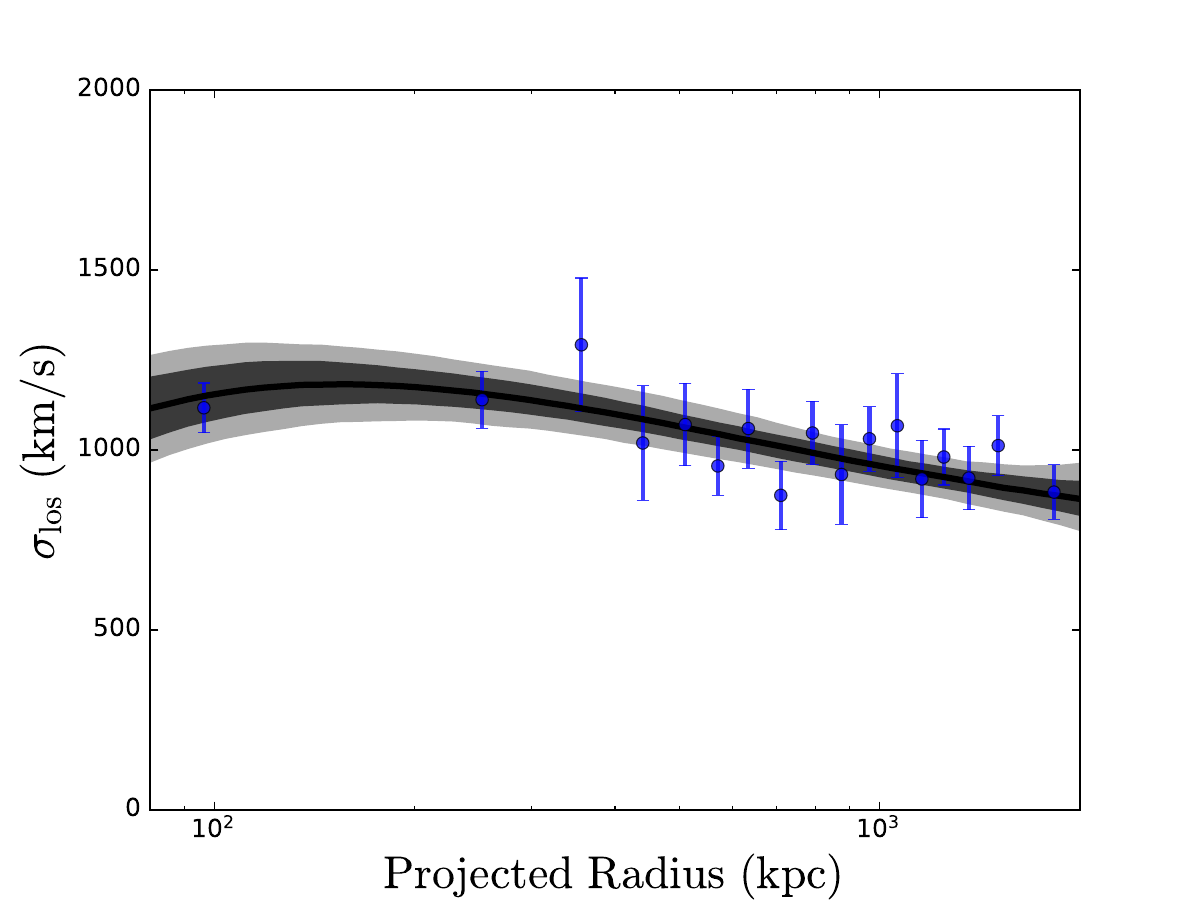}
    \caption{Example fits of projected galaxy number density profiles and line-of-sight velocity dispersion profiles. The cluster shown here is A0085. The galaxy surface number density profile is fitted with three Plummer spheres, which are then used as input in the projected Jeans equation. The line-of-sight velocity dispersion profile is used to determine the total mass profile, parameterized using the cNFWt profile with six parameters. Dark and light shadow regions show the 1$\sigma$ and 2$\sigma$ confidence intervals, respectively.}
    \label{fig:sigma}
\end{figure*}

In this paper, we use the GravSphere code by \citet{Read2017}, which was written to study the dynamics of star clusters and dwarf galaxies using individual stars as dynamical tracers. Since it solves the same equations (eq. \ref{eq:SphereJeans} and \ref{eq:sigmalos}), we can transplant it to galaxy clusters by using individual galaxies as dynamical tracers and changing the corresponding parameters and priors. GravSphere has been carefully validated on a wide array of mock data, including  spherical systems \citep{Read2017, Read2021}, systems with limited spectroscopic data \citep{Collins2021}, triaxial systems \citep{Read2017}, tidally disrupting systems \citep{Read2018, deLeo2023}, and cosmologically realistic mocks \citep{Genina2020}. GravSphere uses parameterized functions to describe the unknown variables in equation \ref{eq:SphereJeans} and \ref{eq:sigmalos}. To guarantee that the chosen functions can provide good fits for diverse objects, we choose flexible functions with sufficient number of parameters. Galaxy number density is modeled with three Plummer spheres \citep{Plummer1911},
\begin{equation}
    \nu(r) = \sum^3_{i=1}\frac{3N_i}{4\pi a^3_i}\Big(1+\frac{r^2}{a^2_i}\Big)^{-5/2}
\end{equation}
where the values of $N_i$ are chosen to make sure $\int\nu(r){\rm d^3}\vec{r}=1$. We set broad boundaries for the parameters: $10^{-4}<N_i<100$ and 50 kpc $<a_i<$ 2000 kpc. It is possible to include more Plummer spheres and hence more free parameters. However, we found that three Plummer spheres with six parameters are enough for all the considered systems. The values of $(N_i, a_i)$ are determined by fitting the projected number density to observed values, which is analytically given by 
\begin{equation}
    \Sigma_{\rm gal} = \sum^3_{i=1}\frac{N_ia^2_i}{\pi(a^2_i+R^2)^2}.
\end{equation}
An example fit with cluster A0085 is given in the left panel of Figure \ref{fig:sigma}.

The total enclosed mass $M(<r)$ is modeled with the cNFWt profile \citep{Read2018}, which is based on the Navarro-Frenk-White model \citep[NFW,][]{Navarro1996},
\begin{equation}
    \rho_{\rm NFW} = \frac{\rho_s}{\frac{r}{r_s}\Big(1+\frac{r}{r_s}\Big)^2}.
\end{equation}
The cNFWt model modifies the NFW profile at both small and large radii. At small radii, the density profile is described as the cNFW model \citep{Read2016} and its enclosed mass is given by
\begin{equation}
    M_{\rm cNFW}(<r) = M_{\rm NFW}(<r)f^n,
\end{equation}
where $f=\tanh{\Big(\frac{r}{r_c}\Big)}$, and the index $n$ controls how cuspy the halo is. A NFW-like cusp has $n=0$, while a complete core requires a minimum value of $n=1$. The corresponding density profile reads as
\begin{equation}
    \rho_{\rm cNFW}(<r) = \rho_{\rm NFW}f^n + \frac{nf^{n-1}(1-f^2)}{4\pi r^2r_c}M_{\rm NFW}(<r).
\end{equation}
At $r>r_t$, the enclosed mass is given by
\begin{equation}
    M_{\rm cNFWt}(<r) = M_{\rm cNFW}(<r_t) + 4\pi\rho_{\rm cNFW}(r_t)\frac{r_t^3}{3-\delta}\Big[\Big(\frac{r}{r_t}\Big)^{3-\delta}-1\Big],
\end{equation}
where $\delta$ is the outer slope. In total, the cNFWt profile has six parameters, which can accommodate different inner and outer slopes. We set loose boundaries as: $13<\log(M_{200}/M_\odot)<20$, $0.1<c_{200}<100$, 10 kpc $<r_c<$ 3000 kpc, $-0.5<n<2.0$, 10 kpc $<r_t<$ 5000 kpc and $0<\delta<3$. The total mass $M_{200}$ of galaxy clusters generally ranges from 10$^{14}$ to 3$\times$10$^{15}$ $M_\odot$, and the halo concentration $c_{200}$ is typically between 1 and 10 \citep{Umetsu2016}. Therefore, the allowed ranges are much wider than the expected values. We allow $r_c$ and $r_t$ to vary from galaxy scale to beyond cluster scale. The core-cusp control parameter $n$ is typically within (0, 1). The additional outer slope parameter $\delta$ is introduced to accommodate density profiles less steep than the NFW model, which can be switched off by setting a large value of $r_t$ if necessary.

By default, the output of the GravSphere code is the enclosed dark matter halo mass, which is calculated by subtracting baryonic mass from the total mass. Since we will have to model stellar mass and gas mass distributions separately, we simply set a negligible baryonic mass in the code. As a result, the output mass profile, as described by the cNFWt model, is the total enclosed mass. The cNFWt model has sufficient freedom to accommodate various mass profiles (see an example fit in the right panel of Figure \ref{fig:sigma}). This is a critical reason why we do not use the NFW model, which is generally believed to work well in galaxy clusters. Therefore, we treat the cNFWt model simply as a fitting function rather than a halo model, and set loose boundaries for its parameters without imposing cosmological priors.

The velocity anisotropy parameter $\beta(r)$ is also modelled by a fitting function to avoid irregular behavior,
\begin{equation}
    \beta(r) = \beta_0 + (\beta_\infty-\beta_0)\frac{1}{1+\big(\frac{r_0}{r}\big)^n},
    \label{eq:velani}
\end{equation}
where $\beta_0$ and $\beta_\infty$ are the values at $r=0$ and $r=\infty$, respectively; $r_0$ and $n$ are used to characterize the radial shape. The boundary for $n$ is set as: $1<n<3$; while $r_0$ is allowed to vary within $0.5R_{\rm half}<r_0<2R_{\rm half}$, where $R_{\rm half}$ is the radius enclosing 50\% of the total galaxies. In order to avoid the infinite value for a full tangential velocity dispersion, \citet{Read2006} defines a symmetrized anisotropy parameter,
\begin{equation}
    \tilde{\beta} = \frac{\beta}{2-\beta}.
\end{equation}
According to this definition, the full tangential and radial velocity dispersion corresponds to $\tilde{\beta}=-1$ and $\tilde{\beta}=1$, respectively. Since the velocity distribution is expected to be nearly isotropic in the center of galaxy clusters due to the strong gravitational field, we set the boundary as $-0.2<\tilde{\beta}_0<0.2$ in the center, while we allow for a larger anisotropy at large radii by setting $-0.2<\tilde{\beta}_\infty<1.0$. We will find the set ranges are sufficiently large for all the considered clusters. 

Notoriously, the velocity anisotropy parameter is degenerated with the density profile, as a larger anisotropy would lead to a larger radial velocity dispersion, which is positively correlated with the enclosed mass profile. The degeneracy can be ameliorated by proper motions, which provide two additional projected equations, similar to eq. \ref{eq:sigmalos}. However, proper motions are hardly measurable for cluster galaxies. \citet{Merrifield1990} found that the fourth order of the velocity distribution can be related to the enclosed mass profiles in two separate and independent ways, i.e.
\begin{eqnarray}
    v_{s1} &=&\frac{2}{5}\int_0^\infty GM\nu(5-2\beta)\sigma^2_rr{\rm d}r\nonumber\\
    &=& \int_0^\infty\Sigma_{\rm gal}\langle v^4_{\rm los}\rangle R{\rm d}R,
\end{eqnarray}
\begin{eqnarray}
    v_{s2} &=&\frac{4}{35}\int_0^\infty GM\nu(7-6\beta)\sigma^2_rr^3{\rm d}r\nonumber\\
    &=& \int_0^\infty\Sigma_{\rm gal}\langle v^4_{\rm los}\rangle R^3{\rm d}R,
\end{eqnarray}
where $\langle v^4_{\rm los}\rangle=\int v^4_{\rm los}f{\rm d^3}\vec{v}$. Therefore, the two Virial Shape parameters can help ameliorate the degeneracy. 

In total, GravSphere solves the Jeans equation with ten parameters (six in cNFWt and four in $\beta(r)$). We impose flat priors on these fitting parameters within the aforementioned hard boundaries. The parameter space is explored using the Markov Chain Monte Carlo method with the $emcee$ hammer by \citet{Foreman-Mackey2013}. We use 250 walkers and run 50 thousand steps. Raising the number of steps to 100 thousand does not affect our results, demonstrating that our chains are converged. In Appendix \ref{sec:corner}, we show some example corner plots for A0085, projected into the space of parameters that are well-constrained by our models.

\begin{figure*}
    \centering
    \includegraphics[scale=0.45]{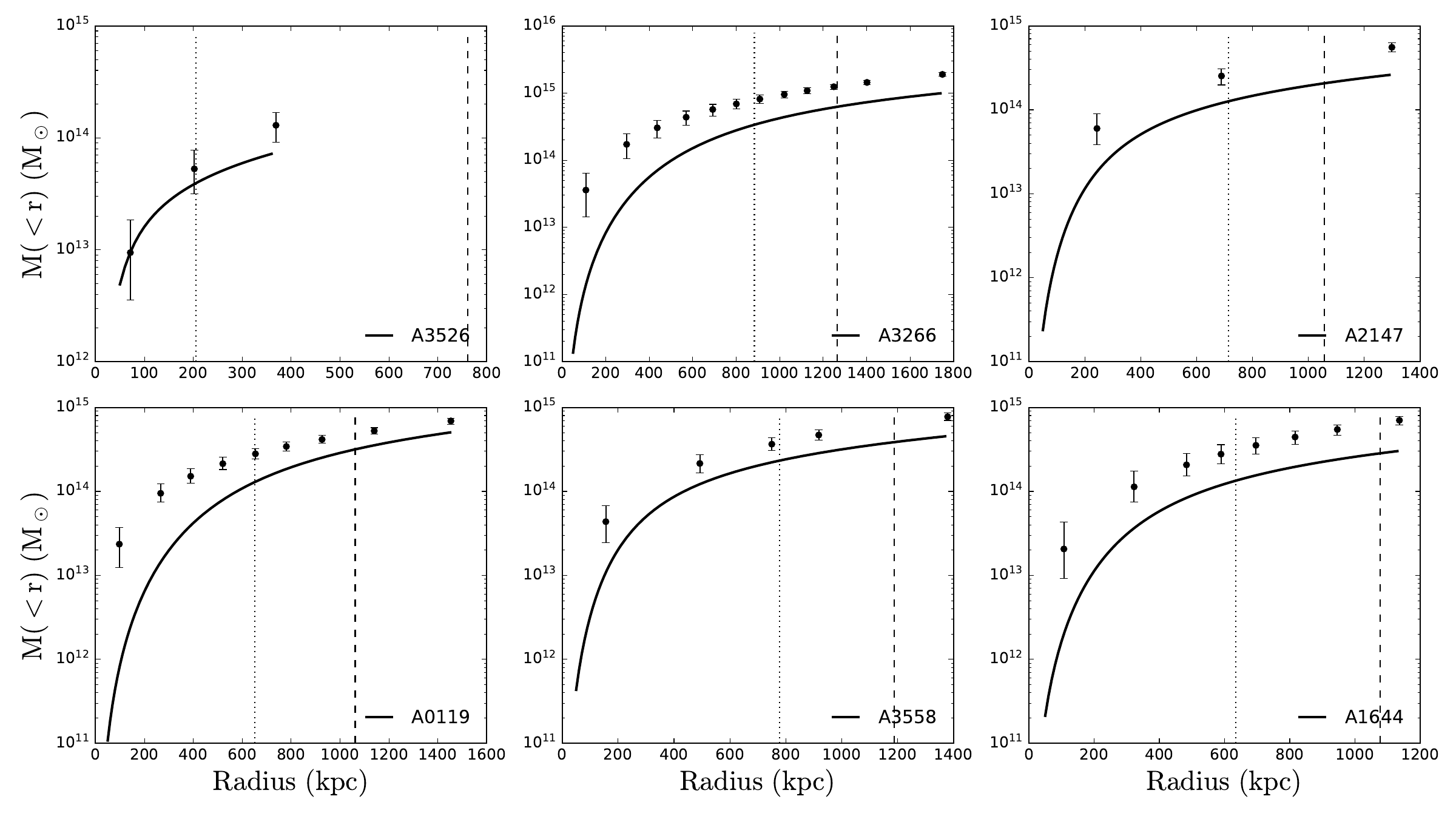}
    \caption{Dynamical mass profiles of six disturbed clusters of galaxies. Solid lines are the hydrostatic mass profiles derived from the surface brightness fits using the $\beta$ function from \citet{Chen2007}. Points with errorbar show the total mass profiles from the line-of-sight velocity dispersion. Vertical dashed lines indicate the positions of $r_{500}$ measured from X-ray data by \citet{Zhang2011}, while dotted lines mark $R_{\rm half}$ enclosing 50\% of the total cluster galaxies. 3D radii are chosen according to the binned projected radii to avoid oversampling. That the dynamical mass profiles determined from different tracers are inconsistent presumably indicates non-equilibrium conditions stemming from recent or ongoing mergers}.
    \label{fig:Mdisturb}
\end{figure*}

\subsection{Binning data with Binulator}

The robust estimate of the dynamical mass is contingent on the accurate calculation of the velocity dispersion, which typically requires a large number of tracers in each bin. This is difficult to achieve in many cases. As such, \citet{Collins2021} introduced a separate routine for data binning, Binulator, prior to running GravSphere. Instead of directly calculating velocity dispersion from measured line-of-sight velocities, Binulator fits a generalized Gaussian function within each bin and estimates the mean, variance and kurtosis. The generalized function is given by
\begin{equation}
    p_i = \frac{\beta}{2\alpha\Gamma(1/\beta)}\exp{\big[-\big(|v_{\rm los,i}-\mu|/\alpha\big)^\beta\big]},
\end{equation}
where $\alpha$, $\beta$, $\mu$ are fitting parameters, the index $i$ labels each individual galaxy, and $\Gamma(x)$ is the Gamma function. As before, we set loose boundaries for these parameters: 50 km/s $<\alpha<$ 2000 km/s, $1<\beta<10$, $|\mu|<$ 1000 km/s. Based on the fit, the velocity dispersion is given by $\sigma^2_{\rm los}=\alpha^2\Gamma(3/\beta)/\Gamma(1/\beta)$. Since there is no error measurement for line-of-sight velocity in the available catalogs, we assume a 10\% uncertainty on the velocity for all cluster galaxies. This is a conservative assumption, since galaxy velocities are measured via spectroscopy with high accuracy. This uncertainty generates an error probability function, which is convolved with the generalized Gaussian function. Eventually, the parameters are determined for each bin by maximizing the likelihood function (with the $emcee$ hammer),
\begin{eqnarray}
    \mathcal{L} = \prod_{i=1}^Np_i,
\end{eqnarray}
where $N$ is the total number of galaxies in each bin. \citet{Collins2021} showed that the Binulator routine can robustly estimate velocity dispersion with 25 galaxies/bin. We hence set the minimum bin size as 25 galaxies, while for rich clusters we can also choose to use a larger bin size.

\section{Disturbed clusters}

Galaxy kinematics traces the gravitational potential only if the cluster is relaxed. This is not always the case, as clusters could be undergoing merging and thereby the distribution and motions of galaxies are disturbed. Unrelaxed clusters can display characteristic signatures in their X-ray images. \citet{Nagai2007, Ventimiglia2008} examined some simulated clusters and classified relaxed and unrelaxed clusters based on the morphology of the mock X-ray images. \citet{Vikhlinin2009} applied their procedure to observed clusters, and identified unrelaxed clusters as those with secondary maxima, filamentary X-ray structures, or significant isophotal centroid shifts in their X-ray images. These clusters may deviate from hydrostatic equilibrium, as they present a systematic offset from the $M_{\rm tot}-T_X$ relation \citep{Mathiesen2001, Kravtsov2006}. \citet{Kravtsov2006} found that the total mass is higher than the expected mass by $17\pm5\%$ at the same temperature. Therefore, \citet{Vikhlinin2009} suggested scaling up the total mass from X-ray temperature by 17\%. However, the actual offset of observed clusters in the $M_{\rm tot}-T_X$ is unclear, which has to be calibrated through weak lensing analysis.

In our sample, six clusters are disturbed according to their X-ray images \citep{Vikhlinin2009, Zhang2011}. We derive their hydrostatic mass from the surface brightness fits using the $\beta$ function (eq. \ref{eq:hydromass}), and plot them in Figure \ref{fig:Mdisturb}. We also plot the mass profiles derived from the line-of-sight velocity dispersion, which are drawn from the best-fit cNFWt function, but we only show the discrete points with errors rather than the full curve. This is intended to avoid over-constraining where there is no actual data. The points are chosen at the projected radii of the binned data, although the 3D and projected radii do not exactly correspond to each other. The errors are estimated from the Markov Chain, so they are formal uncertainties that should be taken with a grain of salt. 

Figure \ref{fig:Mdisturb} shows that the total mass profiles derived from galaxy kinematics for five clusters are above the hydrostatic mass curves at all radii. The differences are larger than 17\%, so even if we scale up the hydrostatic mass as for simulated clusters, the two measurements would not agree. This suggests that the merging process affects cluster galaxies and X-ray gas with different significance. A3526 is the only cluster that presents consistent measurements. Noticeably, its size is significantly smaller than other clusters. This may imply that X-ray gas and satellite galaxies are equally distorted for small clusters, though the sample is too small to make a solid conclusion.

Both measurements of the total mass should not be thought reliable given that their X-ray images present signatures of un-relaxation. We include these clusters in the paper to illustrate how unrelaxed clusters may look in their dynamical mass profiles. We will exclude them from our dynamical tests.

\begin{figure*}[htb!]
    \centering
    \includegraphics[scale=0.45]{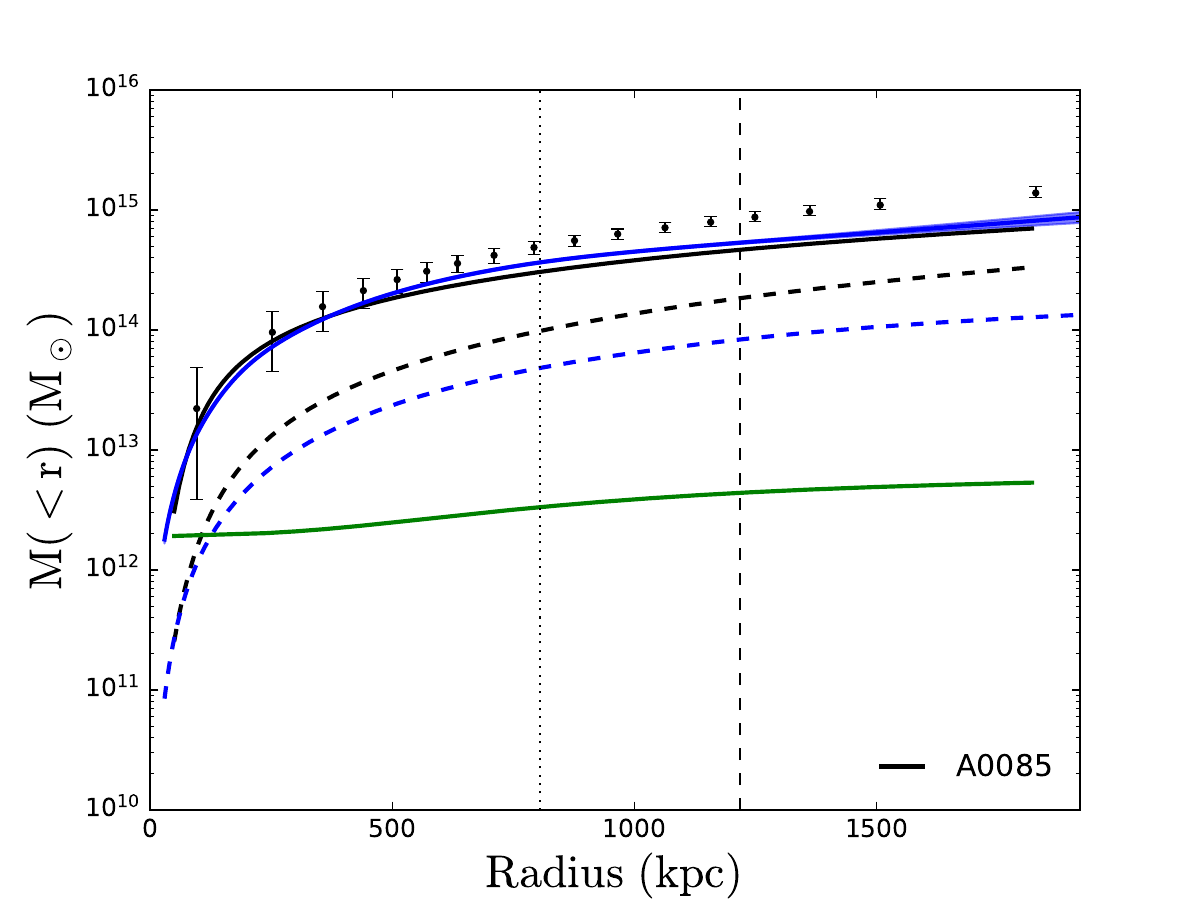}\includegraphics[scale=0.45]{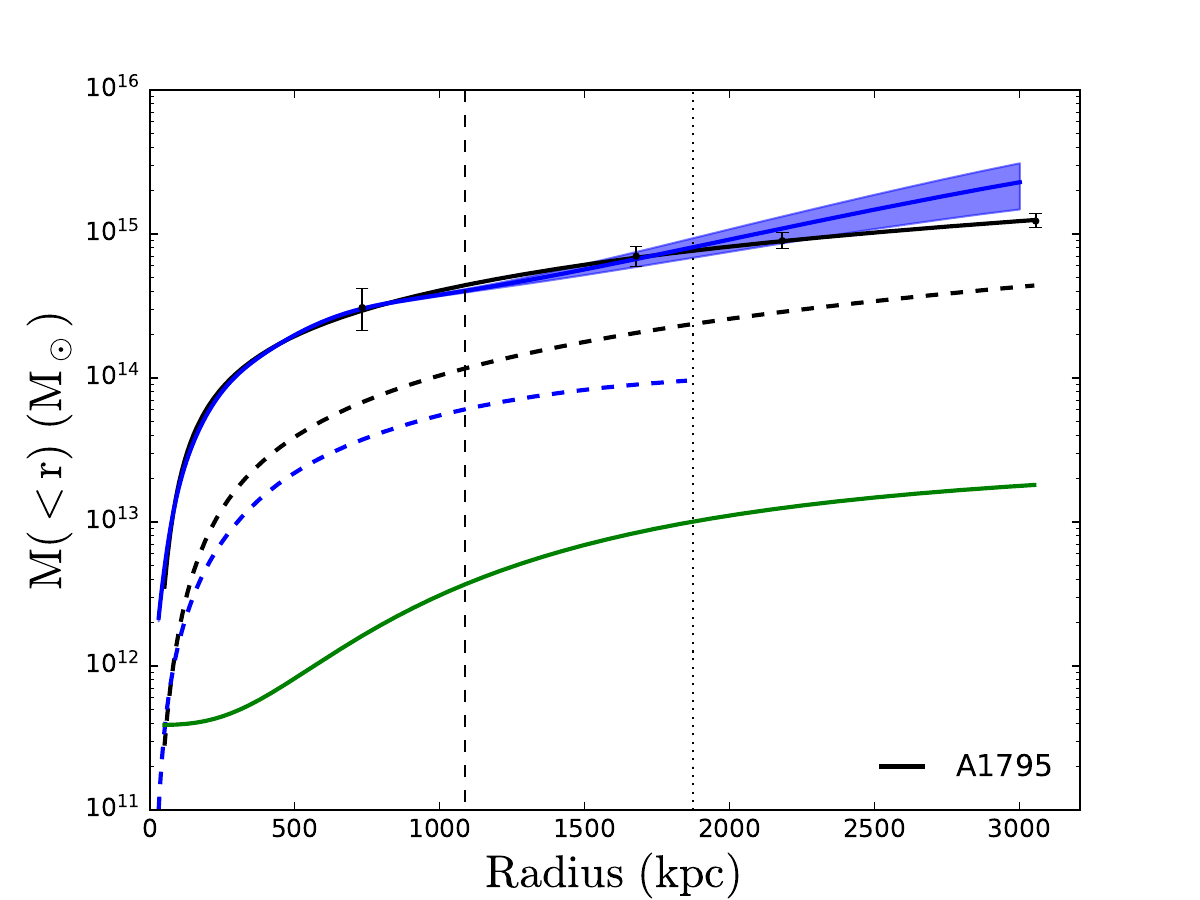}\\
    \includegraphics[scale=0.45]{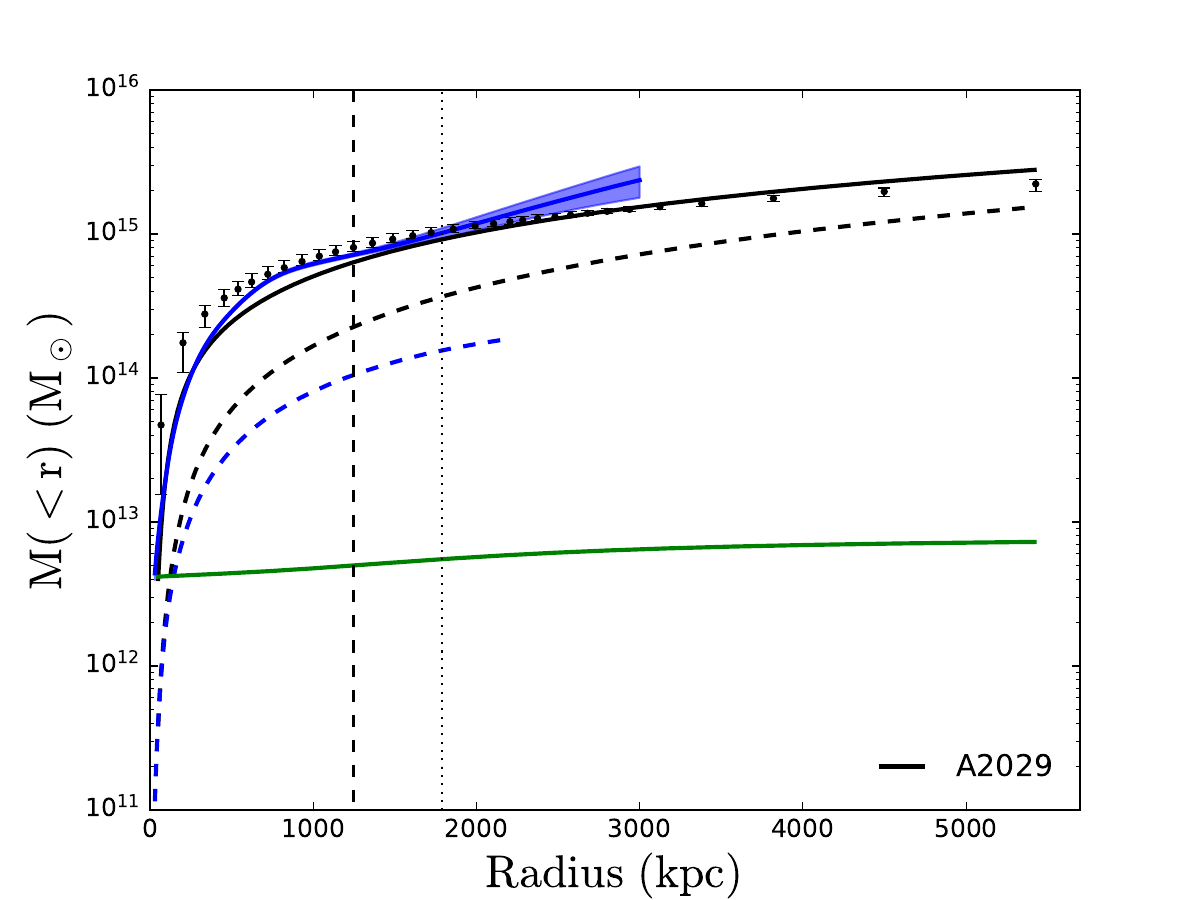}\includegraphics[scale=0.45]{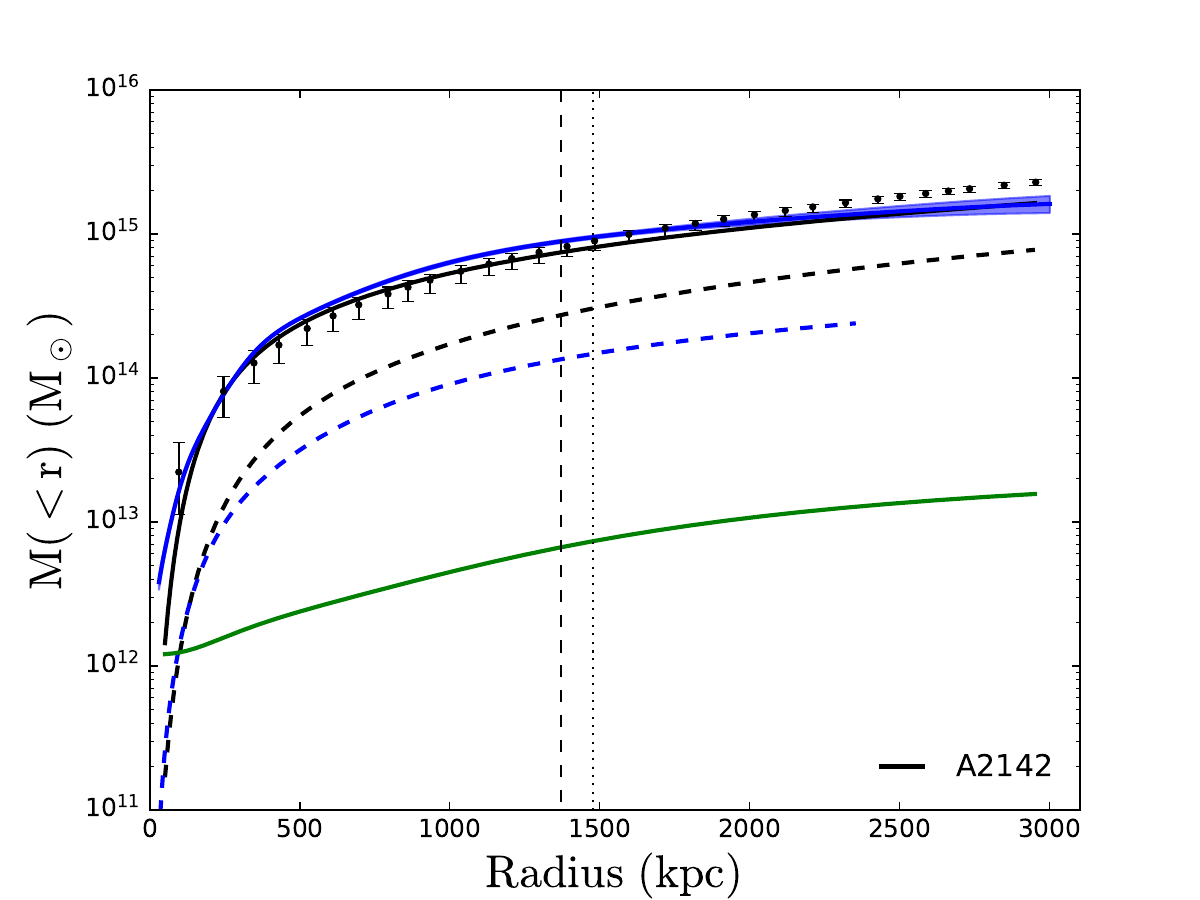}\\
    \includegraphics[scale=0.45]{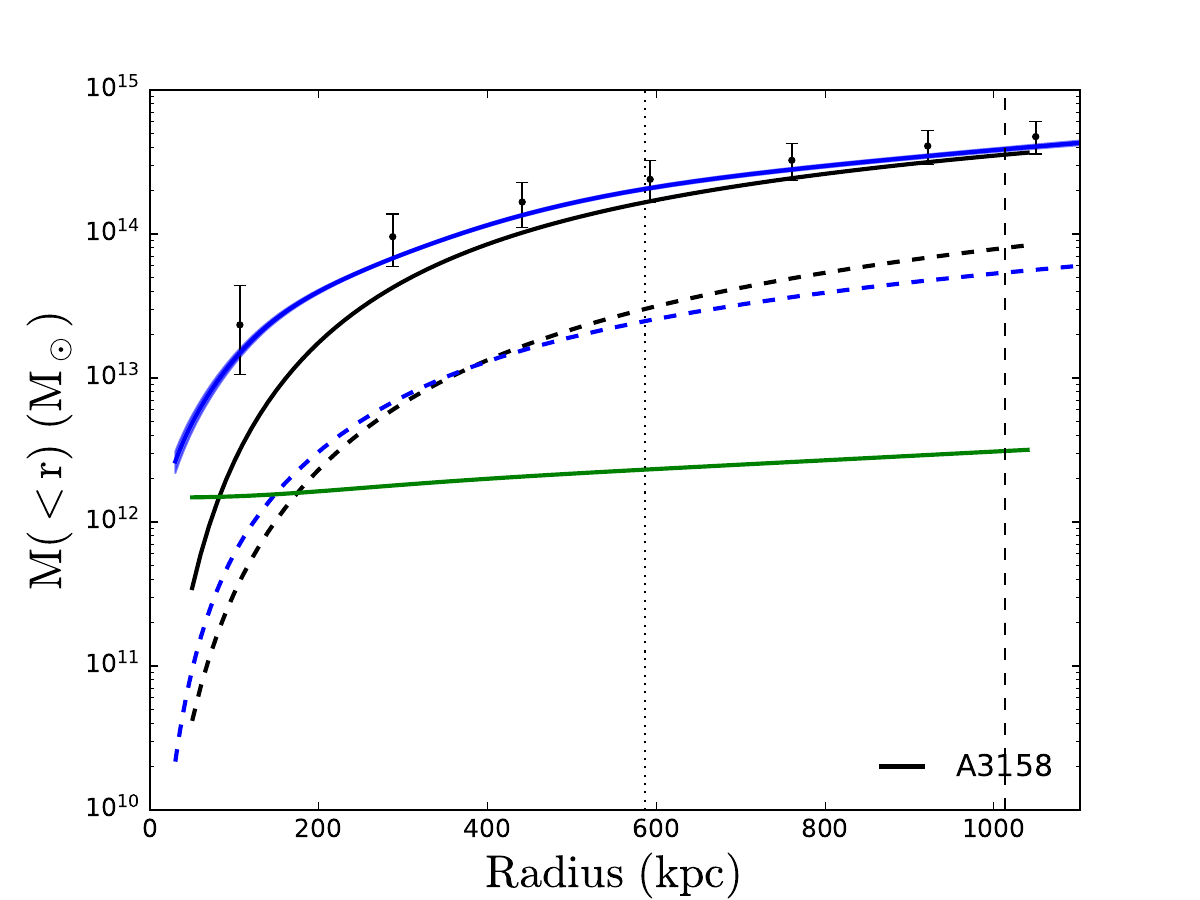}\includegraphics[scale=0.45]{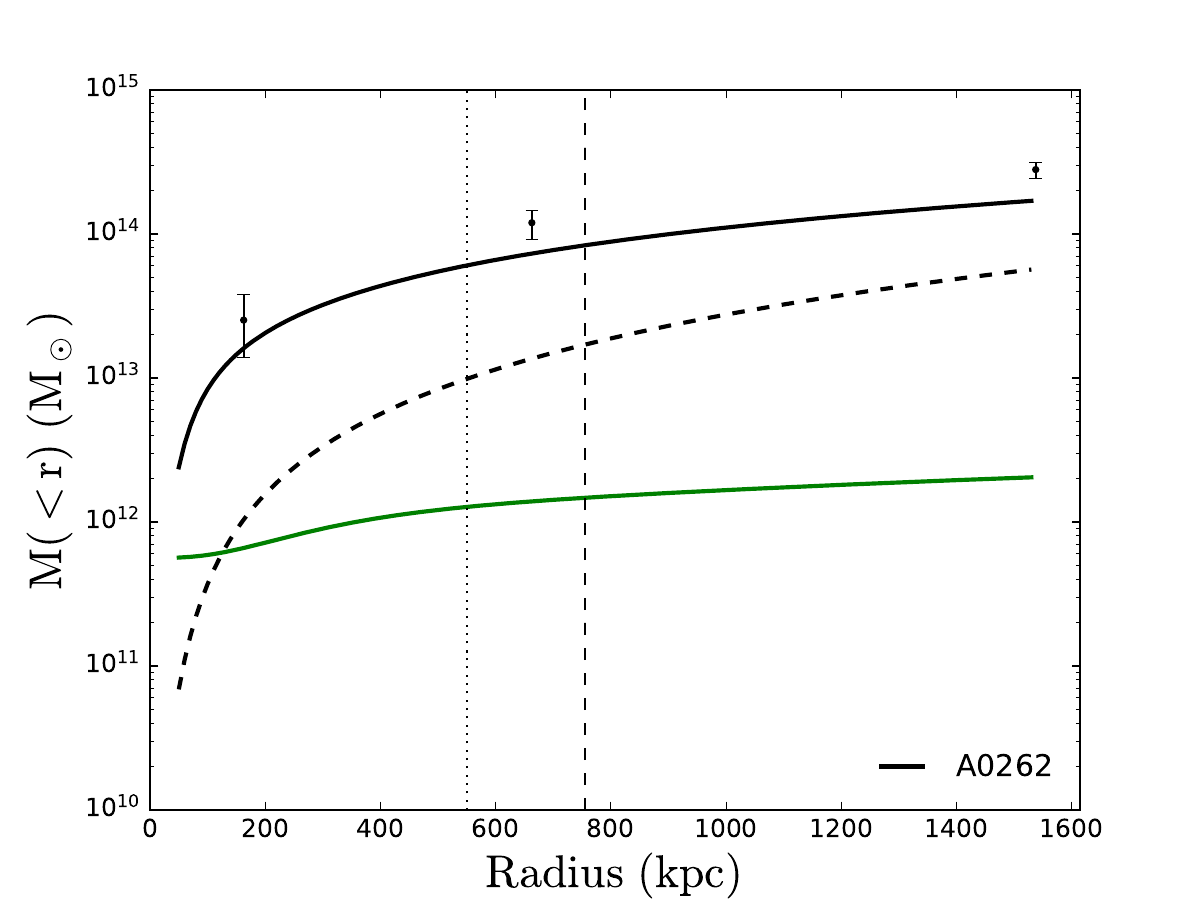}\\
    \caption{Mass profiles of undisturbed clusters of galaxies. Points with errorbar are the total mass profiles from velocity dispersion. Blue solid and dashed lines are the hydrostatic mass and gas mass profiles from the X-COP project, respectively; while black solid and dashed lines show the corresponding results from the surface brightness fits using the $\beta$ functions in \citet{Chen2007}. Green lines show the stellar mass distributions. Vertical dashed lines indicate the positions of $r_{500}$ measured by \citet{Zhang2011} with X-ray data, while dotted lines mark $R_{\rm half}$, which equally divide the total cluster galaxies. For these undisturbed clusters, the dynamical mass profiles from galaxy kinematics are generally consistent with hydrostatic mass profiles, but extend to larger radii in general. Possible reasons for the inconsistency in some clusters are discussed in the text.}
    \label{fig:MassProfile}
\end{figure*}
\renewcommand{\thefigure}{3}
\begin{figure*}
    \centering
    \includegraphics[scale=0.45]{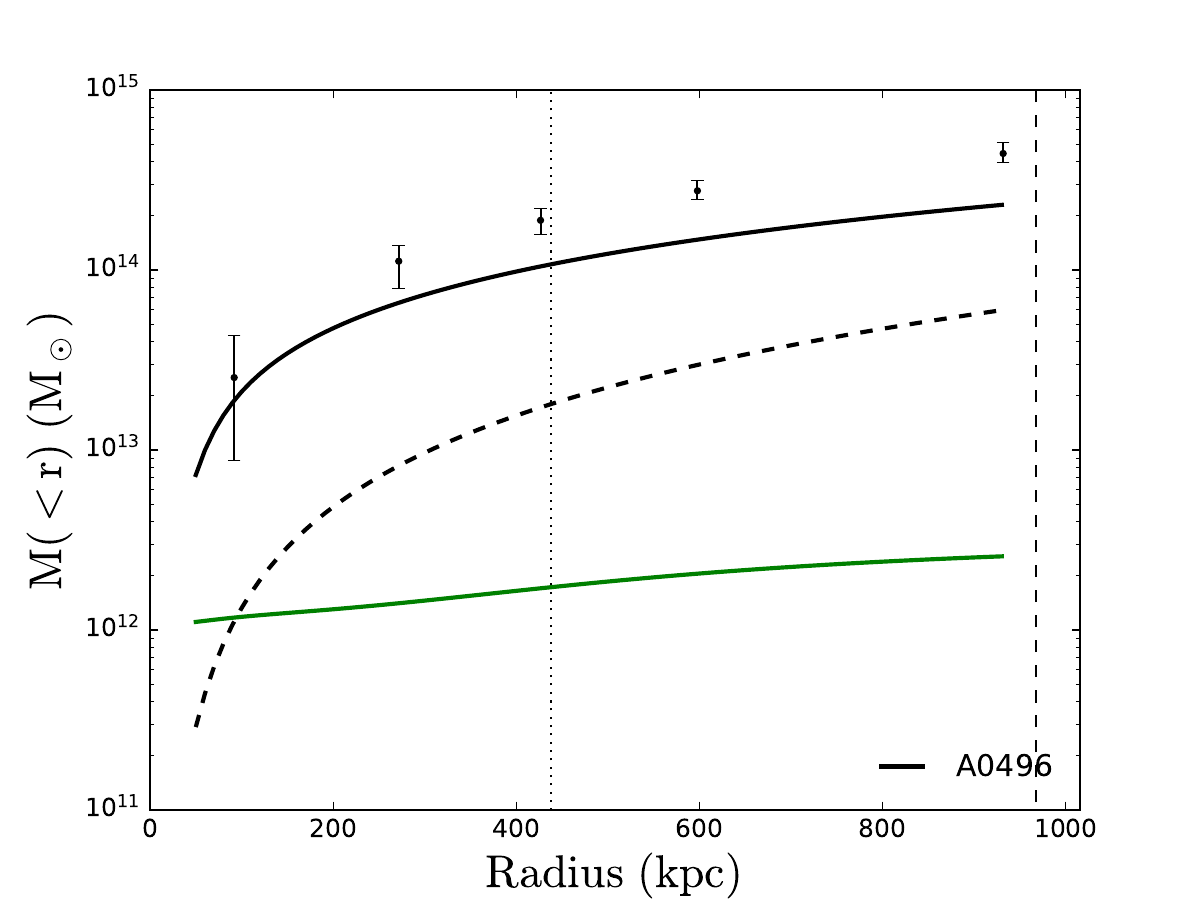}\includegraphics[scale=0.45]{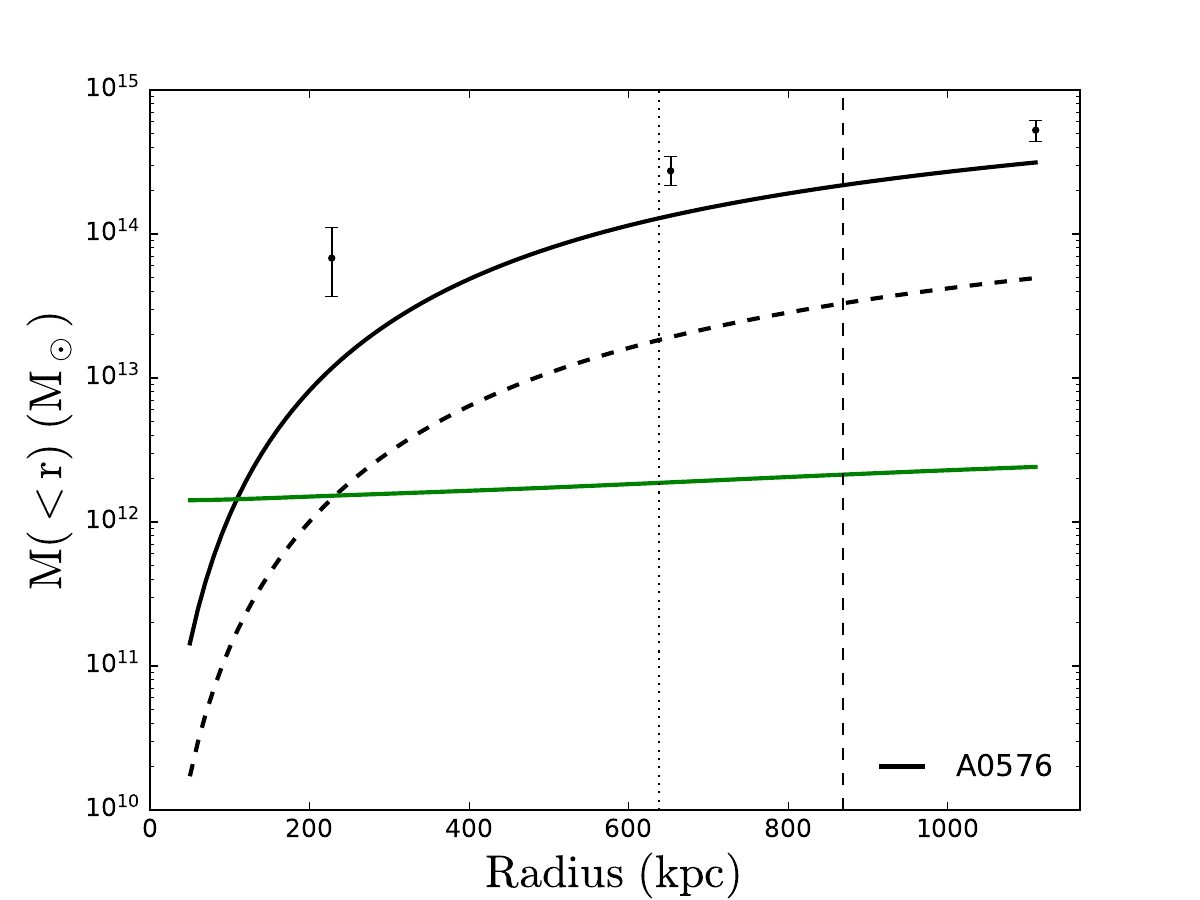}\\
    \includegraphics[scale=0.45]{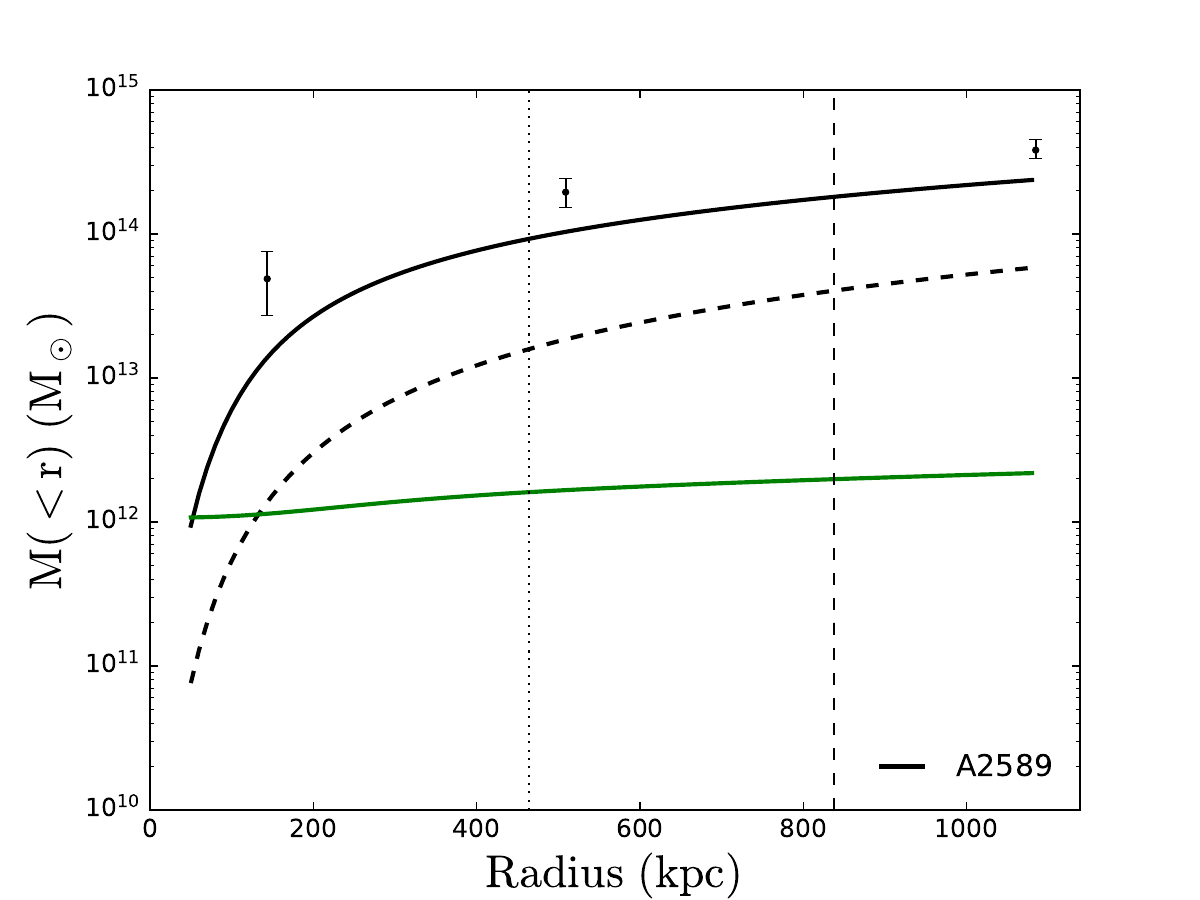}\includegraphics[scale=0.45]{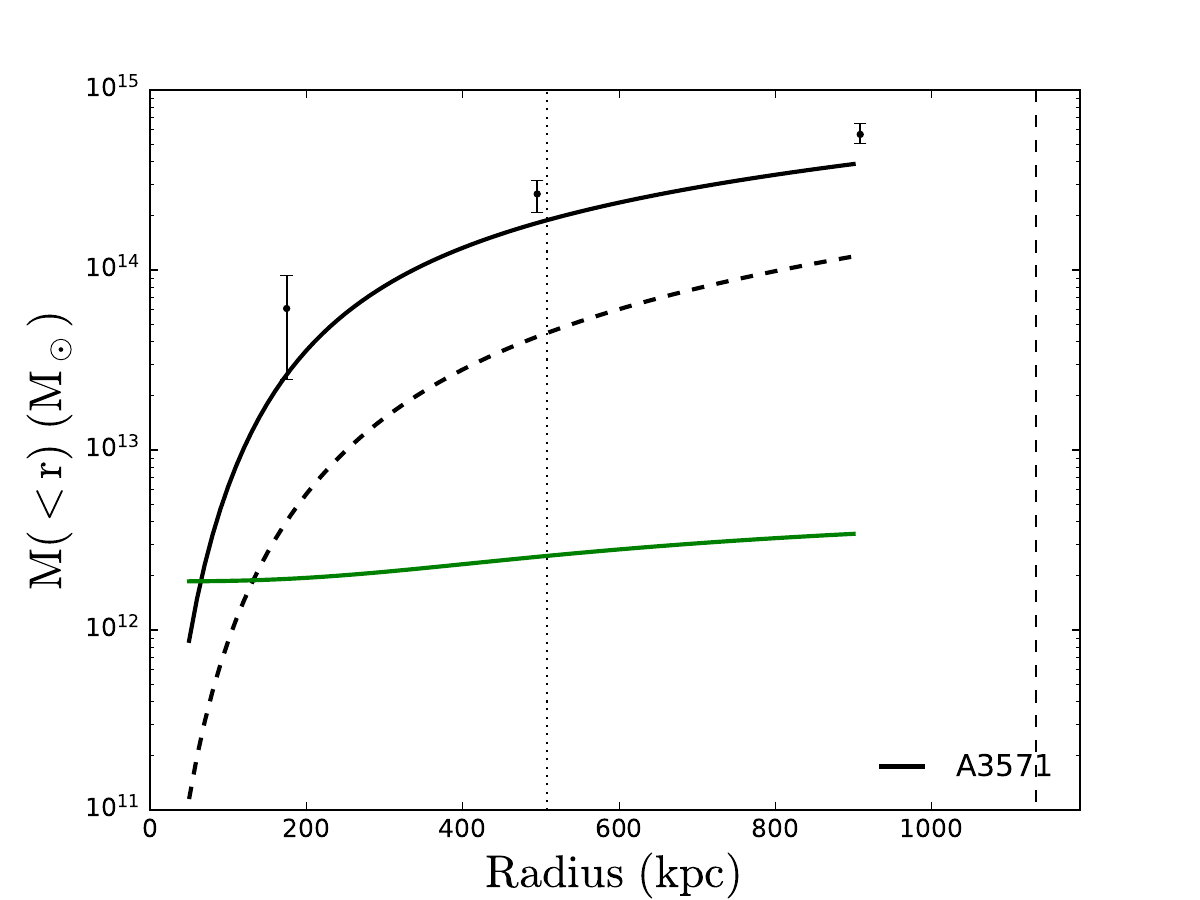}\\
    \caption{continued.}
    \label{fig:Mcontinue}
\end{figure*}

\section{Undisturbed clusters}

\subsection{Baryonic mass profiles}

The ten clusters identified as undisturbed by \citet{Zhang2011} comprise our main sample of interest. We plot their gas mass, stellar mass, hydrostatic mass, and dynamical mass from galaxy kinematics in Figure \ref{fig:MassProfile} and \ref{fig:Mcontinue}. The gas mass profiles are derived by integrating their deprojected surface brightness fits from \citet{Chen2007}. Five clusters have measurements in the XMM Cluster Outskirts Project \citep[X-COP;][]{Eckert2019, Ettori2019, Ghirardini2019}, so we overplot their results for comparison. The gas mass distributions from X-COP are supposed to be more extended those from \citet{Chen2007}, given the aim of the X-COP is to explore the outer regions of clusters and study the growth of structures. Three clusters in our plot present more extended gas profiles from the $\beta$ function fits. These are extrapolations rather than true gas distributions. We extrapolate all the density profiles from the $\beta$ function fits to the outermost radii with the binned galaxy data. We notice that the gas mass from the $\beta$ function is systematically higher than that in X-COP for all five clusters. This is expected given the well-known fact that gas density decreases more quickly than the $\beta$ function predicts at large radii \citep{Vikhlinin2006}. Therefore, the $\beta$ function should be used with caution towards larger radii.

Though gas mass dominates the total baryonic mass in galaxy clusters, the stellar mass contained in galaxies provide significant contributions at small radii. We distinguish BCG and satellite galaxies. Since galaxy kinematics cannot probe the inner regions, we do not need to model the stellar mass distributions of BCGs, but can simply treat them as a point mass. We derive the stellar mass of each BCG based on their $K_s$-band absolute magnitude using the relation from \citet{Cappellari2013ApJL},
\begin{equation}
    \log M_{\rm BCG}/M_\odot \simeq 10.58 - 0.44\times(M_{K_s} + 23),
\end{equation}
where the absolute magnitude is calculated from the observed apparent magnitude from the Two Micron All Sky Survey \citep[2MASS,][]{Skrutskie2006} using the distance measured by Hubble flow. Cappellari's relation corresponds to a stellar mass-to-light ratio between 1.3 and 1.6 $M_\odot/L_\odot$ for our sample, which is higher than the predicted value (1.2 $M_\odot/L_\odot$) of some stellar population models \citep{Schombert2022}.

\begin{table*}[ht!]
	\centering
	\caption{Mass budget of each relaxed cluster. The values of $r_{500}$ and $M_{\rm gas, 500}$ are taken from \citet{Zhang2011}; $M_{\star, 500}$ is calculated using the $M_{\rm gas, 500} - M_{\star, 500}$ relation from \citet{Chiu2018}; $M_{\star,\rm BCG}$ is derived from the $K_s$ band magnitude using the approach by \citet{Cappellari2013}; $M_{\rm dyn,500}$ is interpolated from the best-fit cNFWt mass profile; $M_{\rm hydro,500}$ is recalculated at the same $r_{500}$ to ensure consistency; baryonic fraction is derived based on the dynamical mass $f_b=M_b/M_{\rm dyn,500}$.}
	\label{tab:ClusterMass}
	\begin{tabular}{lccccccccc}
		\hline
		 Cluster & z & $r_{500}$ & $M_{\rm gas, 500}$ & $M_{\star,500}$ &$M_{\star,\rm BCG}$ & $M_{\rm dyn,500}$ & $M_{\rm hydro,500}$ & $M_{\rm dyn,500}$/$M_{\rm hydro,500}$ & $f_b$\\
		     & & (kpc) & ($10^{13}M_\odot$) & ($10^{12}M_\odot$) & ($10^{11}M_\odot$) & ($10^{14}M_\odot$) & ($10^{14}M_\odot$)& &\\
		\hline
A0085 & 0.0554 & 1217 & 6.67 & 4.39 & 19.05 & 8.66 & 4.66 & 1.86 & 0.084\\
A0262 & 0.0162 & 755  & 1.08 & 1.47 & 5.62 & 1.35 & 0.83  & 1.62 & 0.095\\
A0496 & 0.0327 & 967  & 2.79 & 2.60 & 10.72 & 4.65 & 2.39 & 1.94 & 0.068\\
A0576 & 0.0381 & 869  & 2.00 & 2.13 & 14.13 & 4.00 & 2.18 & 1.83 & 0.059\\
A1795 & 0.0613 & 1085 & 4.95 & 3.68 & 3.89 & 4.54 & 4.41  & 1.03 & 0.118\\
A2029 & 0.0779 & 1247 & 8.24 & 4.99 & 41.69 & 8.38 & 6.37 & 1.32 & 0.109\\
A2142 & 0.0908 & 1371 & 13.40& 6.68 & 12.02 & 8.17 & 7.47 & 1.09 & 0.174\\
A2589 & 0.0421 & 837  & 1.77 & 1.98 & 10.72 & 3.11 & 1.81 & 1.72 & 0.067\\
A3158 & 0.0581 & 1013 & 3.75 & 3.11 & 14.79 & 4.56 & 3.55 & 1.28 & 0.092\\
A3571 & 0.0386 & 1133 & 5.16 & 3.77 & 18.62 & 7.47 & 5.02 & 1.49 & 0.077\\
		\hline
	\end{tabular}
\end{table*}

As mentioned in Section 3.2, galaxy number distributions are described by three Plummer spheres $\nu(r)$. For simplicity, we assume that satellite galaxies share the same stellar mass. This should not affect our results as long as the sample of galaxies is sufficiently large. With this assumption, stellar mass distributions are simply galaxy number distributions up to a proportional constant, which can be determined by the total stellar mass. \citet{Chiu2018} reported the correlation between the gas mass and stellar mass within $r_{500}$,
\begin{equation}
    M_\star(<r_{500}) = 4\times10^{12}M_\odot\Big(\frac{M_{\rm gas}(<r_{500})}{5.7\times10^{13}M_\odot}\Big)^{0.6}.
\end{equation}
We take the values of $M_{\rm gas}(<r_{500})$ from \citet{Zhang2011} and derive the enclosed stellar mass. The stellar mass distribution is then given by
\begin{equation}
    M_\star(<r) = (M_\star(<r_{500}) - M_{\rm BCG})\nu(r)/\nu(r_{500}) + M_{\rm BCG},
\end{equation}
which is plotted in green solid lines. Figure \ref{fig:MassProfile} shows the baryonic mass is generally dominated by stellar mass in the inner regions of clusters. 

\subsection{Dynamical mass profiles}

Similar to disturbed clusters, we derive the hydrostatic mass from the surface brightness fits using the $\beta$ function \citep{Chen2007} with a constant temperature. For the five clusters in the X-COP, we overplot their results from \citet{Ettori2019}, which are supposed to be more robust because the radial variations of the temperature are taken into account. \citet{Eckert2022} also included the pressure measurements from the Planck Sunyaev-Zel’dovich (SZ) survey \citep{Planck2014XXIX}, so their dynamical mass profiles are more extended than the gas distributions.

Interestingly, though the gas mass profiles from the X-COP and the $\beta$ function fits differ significantly at large radii, the total mass profiles are quite consistent. This is partially because hydrostatic mass depends on the gas density slope rather than the absolute gas mass density. The slope of the $\beta$ function is shallower than the true density distribution in the outer regions, while the assumed constant temperature over-predicts the actual values. These two deviations have opposite effects on the measurements of hydorstatic mass. As a result, the hydrostatic mass from the $\beta$ function fits with a constant temperature is fairly reliable in spite of the apparent technical caveats. 

In the inner regions, some clusters have a cool core within 0.1$R_{500}$. The positive temperature gradient leads to a smaller pressure gradient, and thereby the derived hydrostatic mass is smaller than that from a constant temperature. On the other hand, the $\beta$ function is known to underestimate the inner slope of the electron number density. The two opposite effects are competing. As a result, the three cool-core clusters, A0085, A1795 and A2029 present roughly consistent measurements with constant and varying temperatures. A2142 is identified as a non-cool-core cluster by \citet{Zhang2011}, but \citet{Eckert2022} show the core temperature is lower than the peak value by ~25\%. The only non-cool-core cluster, A3158, shows that the hydrostatic mass from the $\beta$ function fit is apparently lower than that in X-COP at $r<200$ kpc, which might be due to the shallower inner slope of the $\beta$ function. This does not affect the remaining five clusters for which there are no X-COP results, since we mostly focus on the regions beyond 200 kpc.

In Figure \ref{fig:MassProfile} and \ref{fig:Mcontinue}, we also plot the dynamical mass profiles measured from the line-of-sight velocity dispersion as discrete points with errors. The corresponding total density profiles are presented in the appendix (Figure \ref{fig:DensityProfile} and \ref{fig:Densitycontinue}). Among those five clusters in X-COP, four clusters present roughly consistent mass measurements. These are rich and massive clusters, so their spatial dynamics can be well resolved by galaxy kinematics. The galaxy distribution of A2029 is much more extended than its gas distribution. As such, we can only compare our dynamical mass with the extrapolated hydrostatic mass from the $\beta$ function fit at large radii. The good agreement suggests that the $\beta$ function extrapolation might be a satisfactory approximation. 

The galaxy kinematics of A0085 present consistent dynamical mass within $r_{500}$, while beyond this radius, they overshoot the hydrostatic mass. This indicates that although the X-ray gas is relaxed, the cluster galaxies in the outskirts are not. It could be that relaxation is quicker at small radii but takes longer at large radii. So after merging, the outskirt galaxies need more time to reach equilibrium. One might think that A0085 is growing its structure by accreting more galaxies. However, if these outskirt galaxies have higher velocity dispersion than allowed by the total gravitational potential, they cannot be gravitational bounded within the cluster. Therefore, it is more likely that the hydrostatic mass is underestimated at large radii.

The remaining five clusters are low-mass and low-richness. Figures \ref{fig:MassProfile} and \ref{fig:Mcontinue} show that the dynamical masses from galaxy kinematics for four clusters overshoot their hydrostatic masses, while only A3571 presents an acceptable agreement. The inconsistency may imply that these cluster galaxies are away from dynamical equilibrium, though their X-ray images do not present abnormal structures. If true, this would suggest that galaxies and intracluster medium have different time scales for relaxation. However, our results are qualitatively consistent with what \citet{Foex2017} found. This indicates that their hydrostatic masses are likely being underestimated. 

In Table \ref{tab:ClusterMass}, we present the mass budget for each relaxed cluster. We also compared the total masses measured from galaxy kinematics and hydrostatic equilibrium. Our results show that the total mass from galaxy kinematics is higher than hydrostatic mass by 3\% to 94\%. The mean percentage is $\sim$ 50\%. This could affect the $\sigma_8$ tension \citep{Blanchard2021}. The $\sigma_8$ parameter describes the magnitude of matter fluctuations in the later Universe, which can be calculated from the CMB fluctuations via extrapolations \citep{Planck2016XIII}. It can also be directly measured through cluster counts. The cluster sample for cosmological interests is SZ selected, but the cluster mass is calibrated using hydrostatic mass \citep{Planck2016XXIV}. So if the hydrostatic mass is underestimated, the measured $\sigma_8$ would be lower as well. \citet{Blanchard2021} pointed out the tension can be resolved if the hydrostatic mass is scaled up by 40\%, which is roughly consistent with our mass estimates. However, our sample is quite small, so that it is unclear if the true mass deviation measured from galaxy kinematics and hydrostatic equilibrium is around 40\% or not. To completely resolve the $\sigma_8$ tension, one would also need to understand the cosmic shear analysis from weak lensing measurements \citep{Asgari2021}, which also presents this tension with the CMB results. Our results could be a step forward if statistically confirmed with larger cluster samples.

Though promising for the $\sigma_8$ tension, our measurements result in the baryonic fraction that is in tension with the cosmic value, $f_b=0.16$. Table \ref{tab:ClusterMass} shows that most of our clusters have a baryonic fraction much smaller than 0.16. Therefore, the tension between the early-Universe and late-Universe measurements is hardly resolved in the $\Lambda$CDM cosmology.
\renewcommand{\thefigure}{4}
\begin{figure}
    \centering
    \includegraphics[scale=0.45]{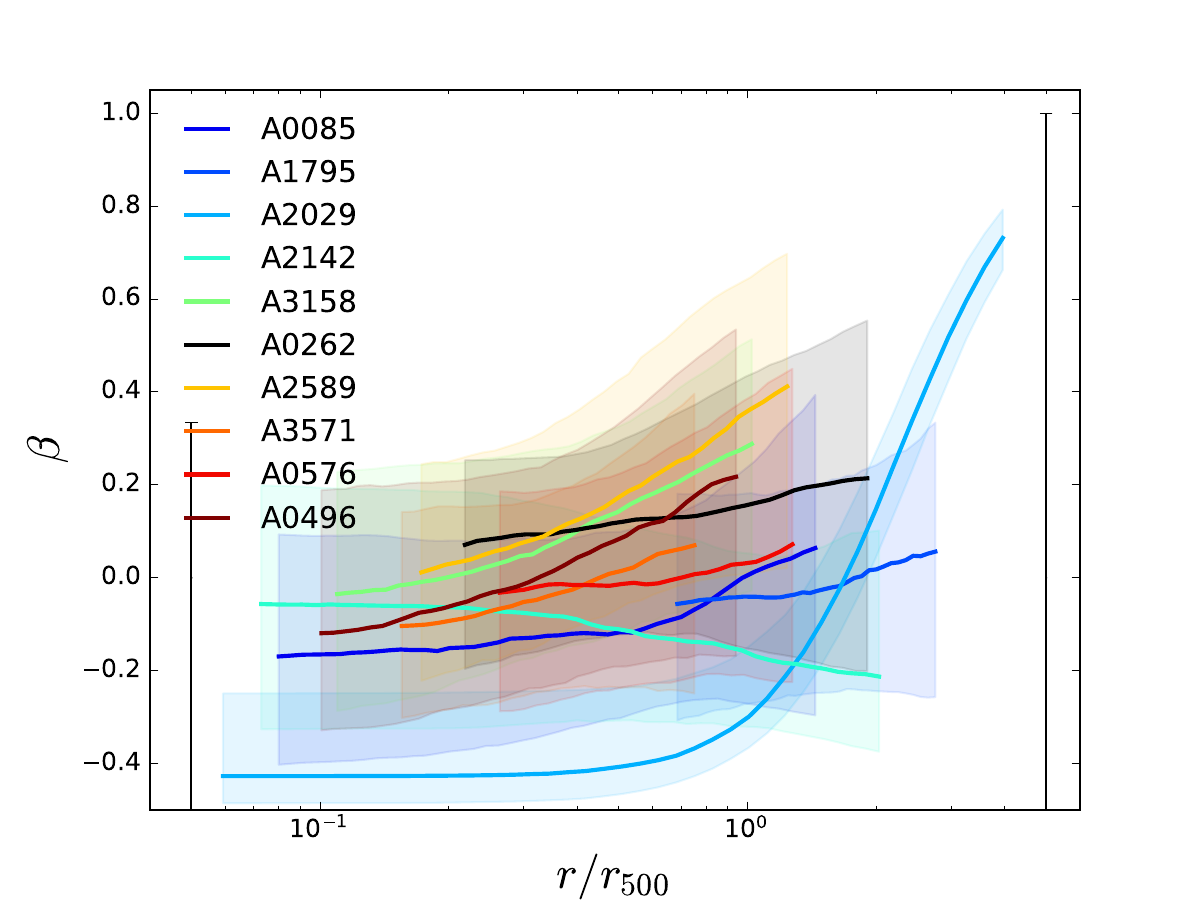}
    \caption{Radial distributions of the velocity anisotropy parameter $\beta$ for the ten undisturbed clusters. Light shadow regions show the 1$\sigma$ confidence intervals. The mean values of $\beta$ are consistent with an isotropic velocity distribution within the errors at small radii, while at large radii the velocity anisotropies are clearly presented in spite of the large errors. The two big error bars at small and large radii represent the allowed ranges set by the priors on $\beta_0$ and $\beta_\infty$, respectively. The lower boundary is $-0.5$ at all radii.}
    \label{fig:beta}
\end{figure}

\subsection{Completeness of the cluster galaxiy sample}

The possible incompleteness of the galaxy samples could affect our mass profiles in two ways: (1) smaller samples make the derived line-of-sight velocity dispersion profiles statistically less robust; (2) it may underestimate the galaxy number density and surface density appearing in equation \ref{eq:SphereJeans} and \ref{eq:sigmalos}. The first concern is largely removed by the Binulator approach, which fits a group of galaxies to the generalized Gaussian function, so missing a few galaxies would not seriously affect the statistics. The second concern is alleviated by the fact that the galaxy number density appears in both the numerator and denominator in equation \ref{eq:SphereJeans} and \ref{eq:sigmalos}, so a constant fraction of incompleteness exactly cancels out (in eq. \ref{eq:sigmalos}, $\Sigma_{\rm gal}$ carries the same incompleteness constant as $\nu(r)$). As a result, only radially variable incompleteness can affect our results. 

Since the clusters in our sample are all at low redshift (z$<$0.1, see Table \ref{tab:ClusterMass}), they have been well observed both photometrically and spectroscopically. Their spectroscopically selected galaxy samples are reported to have a relatively high completeness \citep[e.g. see][]{Cava2009}, so clusters with less extended galaxy distributions are unlikely to have a completeness profile that varies significantly at different radii. For extended clusters like A2142, \citet{Owers2011} showed that its completeness remains almost a constant for up to 3 Mpc (see their Figure 2). Our only concern regarding incompleteness is A2029, the most extended cluster in our sample. \citet{Sohn2017} found that its completeness remains constant up to 1.5 Mpc but decreases by $\sim$15\% at 2 Mpc. In Appendix \ref{sec:incompleteness}, we investigate how the radially varying completeness could possibly bias the mass profile of A2029. We find that even if the completeness decreases by 80\% at the outermost region, the mass profile lies well within the 1$\sigma$ region of that assuming a constant completeness profile. Therefore, incompleteness does not affect our results significantly. Our mass profiles are robust within their errors.

\subsection{Velocity anisotropy}

Figure \ref{fig:beta} plots the radial distributions of the velocity anisotropy parameter for the ten undisturbed clusters. The derivation of their values relies on the two virial shape parameters, $v_{s1}$ and $v_{s2}$, which help ameliorate the $\rho-\beta$ degeneracy. However, $v_{s1}$ and $v_{s2}$ are not radially dependent functions, but two single values. As such, they do not put strong constraints. The measured velocity anisotropies hence present large uncertainties. At small radii, the uncertainties are mostly around 0.1, while they could be as large as 0.2 at large radii. The richest cluster, A2029, has the smallest uncertainty, smaller than 0.1, given that the line-of-sight velocity dispersion profile is better measured thanks to the larger sample of galaxy tracers. 

In spite of the large uncertainties, nine of the ten clusters present similar trends for their mean velocity anisotropies: nearly isotropic in the inner regions while apparently anisotropic in the outskirts. The boundary we impose on the inner anisotropy is $|\tilde{\beta_0}|=|\frac{\beta_0}{2-\beta_0}|<0.2$, which corresponds to $-0.5<\beta_0<0.33$. Figure \ref{fig:beta} shows all the inner anisotropies are well within the set range. This justifies our priors and suggests that the isotropy in the inner regions is truly physical rather than artificially imposed. The only cluster presenting a clear velocity anisotropy in the inner region is A2029, which is an extremely extended cluster. Its stronger constraints most likely owe to its large extent, but it is possible also that they owe to our fitting function (eq. \ref{eq:velani}) being too restrictive. We will explore this in future work.

The apparent anisotropies observed at large radii may not be solid either given their larger uncertainties, except for A2029. Even so, our mean values of velocity anisotropy are qualitatively consistent with the expectations. Since the impact of violent relaxation is strongest at small radii, one expects spherically averaged galaxy motions to present a nearly isotropic velocity distribution \citep[e.g.][]{Pontzen2015}; while at large radii, the velocity distribution could be more radially dominated. Our results are also quantitatively consistent with cosmological simulations \citep{AguirreTagliaferro2021}.

\subsection{Radial acceleration relation}

The extended dynamical mass profiles probed by galaxy kinematics puts strong constraints on dynamical relations. Here we test the radial acceleration relation \citep[RAR,][]{McGaugh2016PRL, OneLaw}, a tight correlation relating the observed acceleration from rotation curves $g_{\rm obs}$ and that from the baryonic distributions $g_{\rm bar}$,
\begin{equation}
    g_{\rm obs} = \frac{g_{\rm bar}}{1-e^{-\sqrt{g_{\rm bar}/g_\dagger}}}.
\end{equation}
This relation was established statistically with 153 late-type galaxies \citep{SPARC} and holds in early-type galaxies \citep{Lelli2017, Shelest2020}. It also holds in individual galaxies once the uncertainties on stellar mass-to-light ratio, galaxy distances and disc inclinations are taken into account \citep{Li2018}. The empirical relation is in line with modified Newtonian dynamics \citep[MOND,][]{Milgrom1983}.
\renewcommand{\thefigure}{5}
\begin{figure}
    \centering
    \includegraphics[scale=0.45]{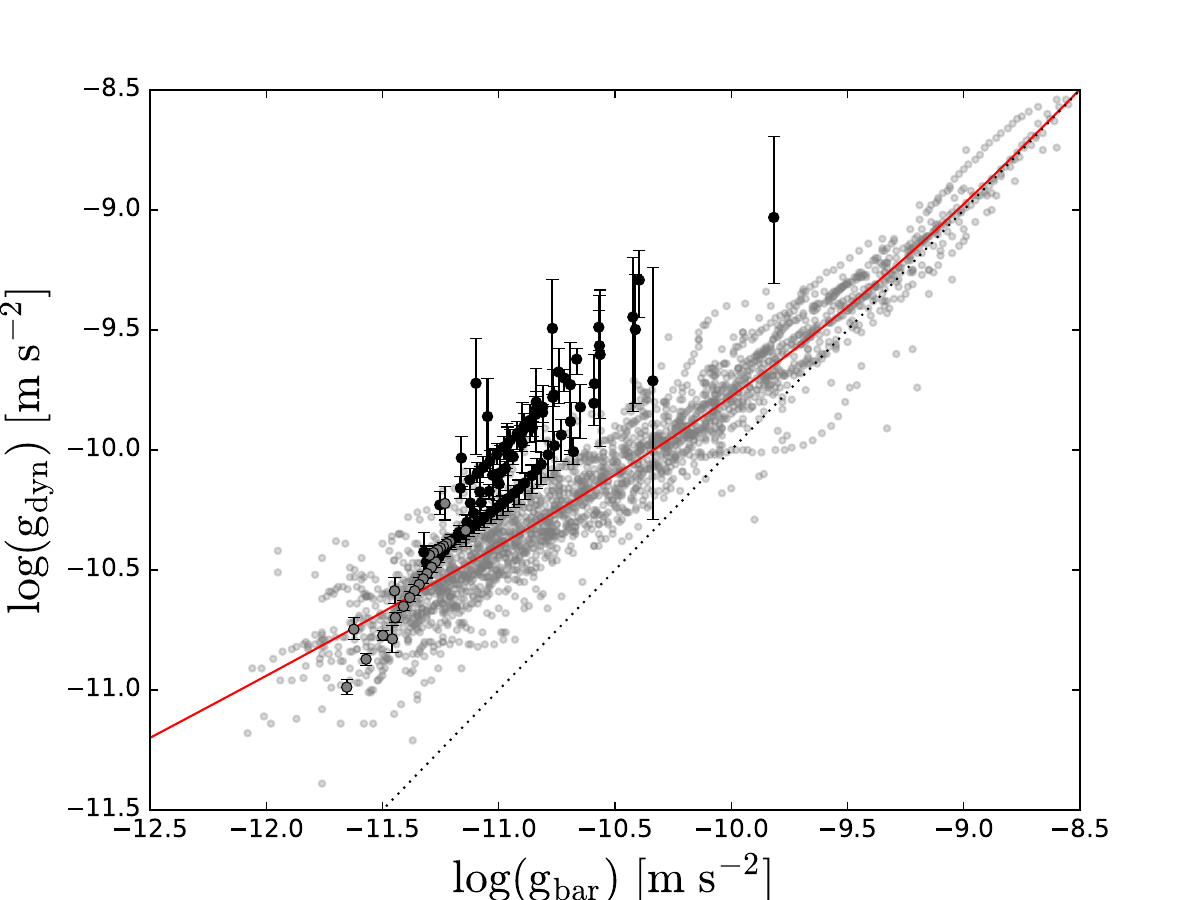}
    \caption{Radial acceleration relation of ten undisturbed galaxy clusters. Black points show the results with both measured dynamical and baryonic masses; while grey points show the radii where there are only measured dynamical masses. The gas masses at these radii are derived from extrapolations based on the surface brightness fits. The dotted line is the line of unity, and the red line is the RAR established with late-type galaxies \citep{McGaugh2016PRL, OneLaw}. Light grey points represent the SPARC galaxies from \citet{McGaugh2016PRL}. The total accelerations overshoot the RAR in galaxies, suggesting a missing baryon problem in the inner regions of galaxy clusters for MOND \citet{Milgrom1983}.}
    \label{fig:RAR}
\end{figure}

The RAR has also been explored on BCG-cluster scales. For example, \citet{Tian2020} employed 20 massive clusters from the Cluster Lensing And Supernova survey with Hubble \citep[CLASH,][]{Postman2012} with total mass profiles measured from gravitational lensing \citep{Umetsu2016}. They found a correlation between $g_{\rm obs}$ and $g_{\rm bar}$ that it is systematically offset from the RAR observed in galaxies, with a significantly larger acceleration scale. \citet{Eckert2022} investigated the RAR using 12 X-COP clusters with dynamical mass profiles measured from hydrostatic equilibrium and the SZ effect. They also reported higher dynamical accelerations than expected from the RAR.

We calculate the total accelerations from the dynamical mass profiles measured with galaxy kinematics,
\begin{equation}
    g_{\rm dyn} = \frac{GM_{\rm dyn}(<r)}{r^2}.
\end{equation}
Since galaxy distributions can be quite extended, their kinematics helps probe lower accelerations than other approaches. The CLASH sample used by \citet{Tian2020} probes accelerations as low as $10^{-10}$ m s$^{-2}$. With the help of the SZ technique, \citet{Eckert2022} extended the acceleration down to slightly above $10^{-11}$ m s$^{-2}$, similar to that observed in individual clusters (Fig.\ \ref{fig:RAR}).

It is challenging to derive the corresponding baryonic acceleration, which is the summation of the gas and stellar contributions,
\begin{equation}
    g_{\rm bar} = \frac{G\big(M_{\rm gas}(<r)+M_\star(<r)\big)}{r^2}.
\end{equation}
In the outskirts, X-ray surface brightness becomes too low to be observable. Therefore, we can only estimate the gas mass via extrapolations using the $\beta$ function fits. For the five clusters in the X-COP, we adopt the modified $\beta$ function from \citet{Vikhlinin2006},
\begin{equation}
    n^2_e = n^2_0\frac{(x/r_c)^\alpha}{(1+x^2/r_c^2)^{3\beta-\alpha/2}}\frac{1}{(1+x^\gamma/r^\gamma_s)^{\epsilon/\gamma}},
\end{equation}
where $x=r/r_{500}$, and $\gamma=3$ is generally fixed. The free parameter space is comprised of ($n_0$, $\alpha$, $\beta$, $\epsilon$, $r_c$, $r_s$). \citet{Ghirardini2019} claimed that the density profile is universal for the 12 X-COP clusters, so they can be described by a single set of values: $\ln n_0=-4.44$, $\alpha=0.89$, $\beta=0.43$, $\epsilon=2.86$, $\ln r_c=-2.99$, $\ln r_s=-0.29$. However, their Figure 3 shows the electron number density profiles are quite diverse in the inner and outer regions. To achieve more accurate extrapolations towards large radii, we only adopt their values of $\alpha$, $\beta$, and $r_c$, but re-fit the outer slope $\epsilon$, scale length $r_s$ and the overall normalization factor $n_0$. The resultant fits properly describe the diversity of the density profiles in the outskirts with greater accuracy, so we extrapolate the gas mass distributions with these fits where X-ray data are absent.

Figure \ref{fig:RAR} plots the total acceleration measured with the line-of-sight velocity dispersions of cluster member galaxies against the baryonic acceleration for the ten relaxed clusters. Regions where the gas mass is estimated by extrapolating the surface brightness fits should be given less credit. Similar to previous works, we find that the total acceleration is systematically higher than expected from the RAR defined by individual galaxies. This is particularly true at intermediate radii. Black points roughly form a line, parallel to the galactic RAR. A similar behavior has been reported by \citet{Tian2020} with the CLASH sample. \citet{Tian2020} found that the higher total acceleration imply a larger critical acceleration ($g_\dagger\sim10^{-9}$ m s$^{-2}$) in clusters than that in galaxies \citep[$g_\dagger\sim10^{-10}$ m s$^{-2}$,][]{McGaugh2016PRL}. The high critical acceleration is a reflection of the missing baryon problem, a well-known problem for MOND in galaxy clusters.

At larger radii and lower acceleration, the data appear to converge with the galactic RAR. This indicates that the missing baryon problem that MOND suffers in clusters is concentrated towards their centers \citep{Sanders2003,Angus2008}, while at large radii the baryonic mass seems consistent with the MOND prediction. However, we caution that the total baryonic mass in the outskirts are extrapolations from surface brightness fits. As a result, the total baryonic mass may be overestimated. Whether the missing baryon problem appears in the outskirts of galaxy clusters remain an open question that will be explored with more extended X-ray data in the future.

\section{Discussion and conclusion}

In this paper, we have applied the spherical Jeans equation to the motions of galaxies in clusters to infer their orbital anisotropies and dynamical mass profiles. We tested this approach with 16 HIFLUGCS clusters and derived the dynamical mass profiles from the observed galaxy distributions and line-of-sight velocity dispersion. We distinguish six clusters with indications of disturbances in their X-ray images, and found their dynamical masses measured from galaxy kinematics are systematically larger than the hydrostatic masses. For the X-ray relaxed clusters, half of them present consistent dynamical mass measurements from galaxy kinematics and hydrostatic equilibrium. These are mostly massive and rich clusters. The rest of the clusters present higher dynamical mass, which suggests that the hydrostatic mass might be underestimated. This could be a step towards solving the $\sigma_8$ tension. The latter requires a higher mass calibration than hydrostatic mass. But it also causes difficulty in reproducing the cosmic baryonic fraction. Alternatively, it could imply that these clusters are not in dynamical equilibrium. Therefore, our approach can also be used to study the dynamical state of galaxy clusters.

A big advantage of using galaxies to constrain cluster dynamics is that they can trace the gravitational potential far beyond X-ray emitting regions, where the acceleration can be extremely low. In this paper, we showed the lowest acceleration probed by the 10 HIFLUGCS clusters is slightly below $10^{-11}$ m s$^{-2}$. The low acceleration region is critical for testing some alternative theories of gravity such as MOND, as well as different dark matter models like superfluid dark matter \citep{Khoury2015}. However, it also raises the concern that galaxies in the outskirts of a cluster may be still on the way of collapsing into the center, so they may have not reached the dynamical equilibrium yet. Since there is no X-ray data in the outskirts, we cannot robustly test the equilibrium, but only provided a rough investigation by extrapolating the hydrostatic mass profiles with the fits using the $\beta$ function or modified $\beta$ function when the actual data are available. The comparison supports the application of galaxy kinematics in the outskirts, which makes our approach promising. A robust calibration can only be carried out by cross-checking the results from weak lensing measurements. Weak lensing measures the projected mass along the line of sight, which includes any foreground or background mass, whereas our dynamical analysis constrains the spherically averaged 3D mass distribution. As such, a comparison between the two requires both a projection of our 3D mass distribution on the sky and analysis of mock data from full cosmological simulations to test the impact of line-of-sight structures. For these reasons, we will consider detailed comparisons with weak lensing studies in future work.

Our dynamical analysis refines and reinforces previous results that MOND suffers a missing baryon problem in rich, $X$-ray emitting galaxy clusters. This discrepancy is pronounced at intermediate radii where the dynamical acceleration exceeds that predicted by the galaxy-calibrated RAR applied to the observed baryons. Numerous solutions have been proposed for this, such as unseen baryons \citep{Milgrom2015}, massive neutrinos \citep{Sanders2003}, sterile neutrinos \citep{Angus2011}, mixed dark matter/alternative gravity models \citep{Berezhiani2015}, making the Lagrangian a function of potential depth as well as acceleration \citep{Hodson2017}, tinkering with the MOND interpolation function \citep{Zhao2012}, or some combination of these. The cluster data converge towards the galaxy RAR at low accelerations in their outskirts. Therefore, it seems the discrepancy is restricted to cluster cores, while the total baryonic mass may be consistent with that predicted by the Baryonic Tully-Fisher relation \citep{McGaugh2005, Lelli2016, McGaugh2021}. However, our baryonic mass estimates in the outskirts are extrapolations and only three clusters extend to the low acceleration region. More clusters with robust gas mass measurements in the outskirts are necessary to further examine this behavior.

\begin{acknowledgements}
We are grateful for the useful discussion with Johan Comparat. P.L. is supported by the Alexander von Humboldt Foundation. Y.T. is supported by Taiwan National Science and Technology Council NSTC 110-2112-M-008-015-MY3. M.S.P. acknowledges funding of a Leibniz-Junior Research Group (project number J94/2020) and a KT Boost Fund by the German Scholars Organization and Klaus Tschira Stiftung. S.S.M. and J.M.S. are supported in part by NASA ADAP grant 80NSSC19k0570. S.S.M. also acknowledges support from NSF PHY-1911909. C.M.K. is supported by Taiwan NSTC 111-2112-M-008-013.
\end{acknowledgements}

\bibliographystyle{aa}
\bibliography{PLi}

\begin{thebibliography}{84}
\expandafter\ifx\csname natexlab\endcsname\relax\def\natexlab#1{#1}\fi

\bibitem[{{Aguirre Tagliaferro} {et~al.}(2021){Aguirre Tagliaferro}, {Biviano},
  {De Lucia}, {Munari}, \& {Garcia Lambas}}]{AguirreTagliaferro2021}
{Aguirre Tagliaferro}, T., {Biviano}, A., {De Lucia}, G., {Munari}, E., \&
  {Garcia Lambas}, D. 2021, \aap, 652, A90

\bibitem[{{Almeida} {et~al.}(2023){Almeida}, {Anderson},
  {Argudo-Fern{\'a}ndez}, {Badenes}, {Barger}, {Barrera-Ballesteros}, {Bender},
  {Benitez}, {Besser}, {Bizyaev}, {Blanton}, {Bochanski}, {Bovy}, {Brandt},
  {Brownstein}, {Buchner}, {Bulbul}, {Burchett}, {Cano D{\'\i}az}, {Carlberg},
  {Casey}, {Chandra}, {Cherinka}, {Chiappini}, {Coker}, {Comparat}, {Conroy},
  {Contardo}, {Cortes}, {Covey}, {Crane}, {Cunha}, {Dabbieri}, {Davidson},
  {Davis}, {De Lee}, {M{\'e}ndez Delgado}, {Demasi}, {Di Mille}, {Donor},
  {Dow}, {Dwelly}, {Eracleous}, {Eriksen}, {Fan}, {Farr}, {Frederick}, {Fries},
  {Frinchaboy}, {Gaensicke}, {Ge}, {Gonz{\'a}lez {\'A}vila}, {Grabowski},
  {Grier}, {Guiglion}, {Gupta}, {Hall}, {Hawkins}, {Hayes}, {Hermes},
  {Hern{\'a}ndez-Garc{\'\i}a}, {Hogg}, {Holtzman}, {Ibarra-Medel}, {Ji},
  {Jofre}, {Johnson}, {Jones}, {Kinemuchi}, {Kluge}, {Koekemoer}, {Kollmeier},
  {Kounkel}, {Krishnarao}, {Krumpe}, {Lacerna}, {Jakson Assuncao Lago},
  {Laporte}, {Liu}, {Liu}, {Liu}, {Lopes}, {Macktoobian}, {Malanushenko},
  {Maoz}, {Masseron}, {Masters}, {Matijevic}, {McBride}, {Medan}, {Merloni},
  {Morrison}, {Myers}, {M{\'e}sz{\'a}ros}, {Negrete}, {Nidever}, {Nitschelm},
  {Oravetz}, {Oravetz}, {Pan}, {Peng}, {Pinsonneault}, {Pogge}, {Qiu},
  {Queiroz}, {Ramirez}, {Rix}, {Fern{\'a}ndez Rosso}, {Runnoe}, {Salvato},
  {Sanchez}, {Santana}, {Saydjari}, {Sayres}, {Schlaufman}, {Schneider},
  {Schwope}, {Serna}, {Shen}, {Sobeck}, {Song}, {Souto}, {Spoo}, {Stassun},
  {Steinmetz}, {Straumit}, {Stringfellow}, {S{\'a}nchez-Gallego},
  {Taghizadeh-Popp}, {Tayar}, {Thakar}, {Tissera}, {Tkachenko}, {Hernandez
  Toledo}, {Trakhtenbrot}, {Fernandez Trincado}, {Troup}, {Trump}, {Tuttle},
  {Ulloa}, {Vazquez-Mata}, {Alfaro}, {Villanova}, {Wachter}, {Weijmans},
  {Wheeler}, {Wilson}, {Wojno}, {Wolf}, {Xue}, {Ybarra}, {Zari}, \&
  {Zasowski}}]{Almeida2023}
{Almeida}, A., {Anderson}, S.~F., {Argudo-Fern{\'a}ndez}, M., {et~al.} 2023,
  arXiv e-prints, arXiv:2301.07688

\bibitem[{{Andernach} {et~al.}(2005){Andernach}, {Tago}, {Einasto}, {Einasto},
  \& {Jaaniste}}]{Andernach2005}
{Andernach}, H., {Tago}, E., {Einasto}, M., {Einasto}, J., \& {Jaaniste}, J.
  2005, in Astronomical Society of the Pacific Conference Series, Vol. 329,
  Nearby Large-Scale Structures and the Zone of Avoidance, ed. A.~P. {Fairall}
  \& P.~A. {Woudt}, 283--287

\bibitem[{{Anders} \& {Grevesse}(1989)}]{Anders1989}
{Anders}, E. \& {Grevesse}, N. 1989, \gca, 53, 197

\bibitem[{{Angus} \& {Diaferio}(2011)}]{Angus2011}
{Angus}, G.~W. \& {Diaferio}, A. 2011, \mnras, 417, 941

\bibitem[{{Angus} {et~al.}(2008){Angus}, {Famaey}, \& {Buote}}]{Angus2008}
{Angus}, G.~W., {Famaey}, B., \& {Buote}, D.~A. 2008, \mnras, 387, 1470

\bibitem[{{Asgari} {et~al.}(2021){Asgari}, {Lin}, {Joachimi}, {Giblin},
  {Heymans}, {Hildebrandt}, {Kannawadi}, {St{\"o}lzner}, {Tr{\"o}ster}, {van
  den Busch}, {Wright}, {Bilicki}, {Blake}, {de Jong}, {Dvornik}, {Erben},
  {Getman}, {Hoekstra}, {K{\"o}hlinger}, {Kuijken}, {Miller}, {Radovich},
  {Schneider}, {Shan}, \& {Valentijn}}]{Asgari2021}
{Asgari}, M., {Lin}, C.-A., {Joachimi}, B., {et~al.} 2021, \aap, 645, A104

\bibitem[{{Berezhiani} \& {Khoury}(2015)}]{Berezhiani2015}
{Berezhiani}, L. \& {Khoury}, J. 2015, \prd, 92, 103510

\bibitem[{{Binney} \& {Mamon}(1982)}]{Binney1982}
{Binney}, J. \& {Mamon}, G.~A. 1982, \mnras, 200, 361

\bibitem[{{Binney} \& {Tremaine}(2008)}]{Binney2008}
{Binney}, J. \& {Tremaine}, S. 2008, {Galactic Dynamics: Second Edition}

\bibitem[{{Biviano} {et~al.}(2006){Biviano}, {Murante}, {Borgani}, {Diaferio},
  {Dolag}, \& {Girardi}}]{Biviano2006}
{Biviano}, A., {Murante}, G., {Borgani}, S., {et~al.} 2006, \aap, 456, 23

\bibitem[{{Biviano} {et~al.}(2013){Biviano}, {Rosati}, {Balestra}, {Mercurio},
  {Girardi}, {Nonino}, {Grillo}, {Scodeggio}, {Lemze}, {Kelson}, {Umetsu},
  {Postman}, {Zitrin}, {Czoske}, {Ettori}, {Fritz}, {Lombardi}, {Maier},
  {Medezinski}, {Mei}, {Presotto}, {Strazzullo}, {Tozzi}, {Ziegler},
  {Annunziatella}, {Bartelmann}, {Benitez}, {Bradley}, {Brescia}, {Broadhurst},
  {Coe}, {Demarco}, {Donahue}, {Ford}, {Gobat}, {Graves}, {Koekemoer},
  {Kuchner}, {Melchior}, {Meneghetti}, {Merten}, {Moustakas}, {Munari},
  {Reg{\H{o}}s}, {Sartoris}, {Seitz}, \& {Zheng}}]{Biviano2013}
{Biviano}, A., {Rosati}, P., {Balestra}, I., {et~al.} 2013, \aap, 558, A1

\bibitem[{{Blanchard} \& {Ili{\'c}}(2021)}]{Blanchard2021}
{Blanchard}, A. \& {Ili{\'c}}, S. 2021, \aap, 656, A75

\bibitem[{{B{\"o}hringer} {et~al.}(2014){B{\"o}hringer}, {Chon}, \&
  {Collins}}]{Bohringer2014}
{B{\"o}hringer}, H., {Chon}, G., \& {Collins}, C.~A. 2014, \aap, 570, A31

\bibitem[{{Cappellari}(2013)}]{Cappellari2013ApJL}
{Cappellari}, M. 2013, \apjl, 778, L2

\bibitem[{{Cappellari} {et~al.}(2013){Cappellari}, {Scott}, {Alatalo}, {Blitz},
  {Bois}, {Bournaud}, {Bureau}, {Crocker}, {Davies}, {Davis}, {de Zeeuw},
  {Duc}, {Emsellem}, {Khochfar}, {Krajnovi{\'c}}, {Kuntschner}, {McDermid},
  {Morganti}, {Naab}, {Oosterloo}, {Sarzi}, {Serra}, {Weijmans}, \&
  {Young}}]{Cappellari2013}
{Cappellari}, M., {Scott}, N., {Alatalo}, K., {et~al.} 2013, \mnras, 432, 1709

\bibitem[{{Cava} {et~al.}(2009){Cava}, {Bettoni}, {Poggianti}, {Couch},
  {Moles}, {Varela}, {Biviano}, {D'Onofrio}, {Dressler}, {Fasano}, {Fritz},
  {Kj{\ae}rgaard}, {Ramella}, \& {Valentinuzzi}}]{Cava2009}
{Cava}, A., {Bettoni}, D., {Poggianti}, B.~M., {et~al.} 2009, \aap, 495, 707

\bibitem[{{Cavaliere} \& {Fusco-Femiano}(1976)}]{Cavaliere1976}
{Cavaliere}, A. \& {Fusco-Femiano}, R. 1976, \aap, 49, 137

\bibitem[{{Chen} {et~al.}(2007){Chen}, {Reiprich}, {B{\"o}hringer}, {Ikebe}, \&
  {Zhang}}]{Chen2007}
{Chen}, Y., {Reiprich}, T.~H., {B{\"o}hringer}, H., {Ikebe}, Y., \& {Zhang},
  Y.~Y. 2007, \aap, 466, 805

\bibitem[{{Chiu} {et~al.}(2018){Chiu}, {Mohr}, {McDonald}, {Bocquet}, {Desai},
  {Klein}, {Israel}, {Ashby}, {Stanford}, {Benson}, {Brodwin}, {Abbott},
  {Abdalla}, {Allam}, {Annis}, {Bayliss}, {Benoit-L{\'e}vy}, {Bertin}, {Bleem},
  {Brooks}, {Buckley-Geer}, {Bulbul}, {Capasso}, {Carlstrom}, {Rosell},
  {Carretero}, {Castander}, {Cunha}, {D'Andrea}, {da Costa}, {Davis}, {Diehl},
  {Dietrich}, {Doel}, {Drlica-Wagner}, {Eifler}, {Evrard}, {Flaugher},
  {Garc{\'\i}a-Bellido}, {Garmire}, {Gaztanaga}, {Gerdes}, {Gonzalez}, {Gruen},
  {Gruendl}, {Gschwend}, {Gupta}, {Gutierrez}, {Hlavacek-L}, {Honscheid},
  {James}, {Jeltema}, {Kraft}, {Krause}, {Kuehn}, {Kuhlmann}, {Kuropatkin},
  {Lahav}, {Lima}, {Maia}, {Marshall}, {Melchior}, {Menanteau}, {Miquel},
  {Murray}, {Nord}, {Ogando}, {Plazas}, {Rapetti}, {Reichardt}, {Romer},
  {Roodman}, {Sanchez}, {Saro}, {Scarpine}, {Schindler}, {Schubnell}, {Sharon},
  {Smith}, {Smith}, {Soares-Santos}, {Sobreira}, {Stalder}, {Stern},
  {Strazzullo}, {Suchyta}, {Swanson}, {Tarle}, {Vikram}, {Walker}, {Weller}, \&
  {Zhang}}]{Chiu2018}
{Chiu}, I., {Mohr}, J.~J., {McDonald}, M., {et~al.} 2018, \mnras, 478, 3072

\bibitem[{{Collins} {et~al.}(2021){Collins}, {Read}, {Ibata}, {Rich}, {Martin},
  {Pe{\~n}arrubia}, {Chapman}, {Tollerud}, \& {Weisz}}]{Collins2021}
{Collins}, M. L.~M., {Read}, J.~I., {Ibata}, R.~A., {et~al.} 2021, \mnras, 505,
  5686

\bibitem[{{De Leo} {et~al.}(2023){De Leo}, {Read}, {Noel}, {Erkal}, {Massana},
  \& {Carrera}}]{deLeo2023}
{De Leo}, M., {Read}, J.~I., {Noel}, N. E.~D., {et~al.} 2023, arXiv e-prints,
  arXiv:2303.08838

\bibitem[{{Diaferio}(1999)}]{Diaferio1999}
{Diaferio}, A. 1999, \mnras, 309, 610

\bibitem[{{Diaferio} \& {Geller}(1997)}]{Diaferio1997}
{Diaferio}, A. \& {Geller}, M.~J. 1997, \apj, 481, 633

\bibitem[{{Eckert} {et~al.}(2022){Eckert}, {Ettori}, {Pointecouteau}, {van der
  Burg}, \& {Loubser}}]{Eckert2022}
{Eckert}, D., {Ettori}, S., {Pointecouteau}, E., {van der Burg}, R.~F.~J., \&
  {Loubser}, S.~I. 2022, \aap, 662, A123

\bibitem[{{Eckert} {et~al.}(2019){Eckert}, {Ghirardini}, {Ettori}, {Rasia},
  {Biffi}, {Pointecouteau}, {Rossetti}, {Molendi}, {Vazza}, {Gastaldello},
  {Gaspari}, {De Grandi}, {Ghizzardi}, {Bourdin}, {Tchernin}, \&
  {Roncarelli}}]{Eckert2019}
{Eckert}, D., {Ghirardini}, V., {Ettori}, S., {et~al.} 2019, \aap, 621, A40

\bibitem[{{Ettori} {et~al.}(2019){Ettori}, {Ghirardini}, {Eckert},
  {Pointecouteau}, {Gastaldello}, {Sereno}, {Gaspari}, {Ghizzardi},
  {Roncarelli}, \& {Rossetti}}]{Ettori2019}
{Ettori}, S., {Ghirardini}, V., {Eckert}, D., {et~al.} 2019, \aap, 621, A39

\bibitem[{{Fo{\"e}x} {et~al.}(2017){Fo{\"e}x}, {B{\"o}hringer}, \&
  {Chon}}]{Foex2017}
{Fo{\"e}x}, G., {B{\"o}hringer}, H., \& {Chon}, G. 2017, \aap, 606, A122

\bibitem[{{Foreman-Mackey} {et~al.}(2013){Foreman-Mackey}, {Hogg}, {Lang}, \&
  {Goodman}}]{Foreman-Mackey2013}
{Foreman-Mackey}, D., {Hogg}, D.~W., {Lang}, D., \& {Goodman}, J. 2013, \pasp,
  125, 306

\bibitem[{{Genina} {et~al.}(2020){Genina}, {Read}, {Frenk}, {Cole},
  {Ben{\'\i}tez-Llambay}, {Ludlow}, {Navarro}, {Oman}, \&
  {Robertson}}]{Genina2020}
{Genina}, A., {Read}, J.~I., {Frenk}, C.~S., {et~al.} 2020, \mnras, 498, 144

\bibitem[{{Ghirardini} {et~al.}(2019){Ghirardini}, {Eckert}, {Ettori},
  {Pointecouteau}, {Molendi}, {Gaspari}, {Rossetti}, {De Grandi}, {Roncarelli},
  {Bourdin}, {Mazzotta}, {Rasia}, \& {Vazza}}]{Ghirardini2019}
{Ghirardini}, V., {Eckert}, D., {Ettori}, S., {et~al.} 2019, \aap, 621, A41

\bibitem[{{Hodson} \& {Zhao}(2017)}]{Hodson2017}
{Hodson}, A.~O. \& {Zhao}, H. 2017, \aap, 598, A127

\bibitem[{{Khoury}(2015)}]{Khoury2015}
{Khoury}, J. 2015, \prd, 91, 024022

\bibitem[{{Kravtsov} {et~al.}(2006){Kravtsov}, {Vikhlinin}, \&
  {Nagai}}]{Kravtsov2006}
{Kravtsov}, A.~V., {Vikhlinin}, A., \& {Nagai}, D. 2006, \apj, 650, 128

\bibitem[{{Lau} {et~al.}(2009){Lau}, {Kravtsov}, \& {Nagai}}]{Lau2009}
{Lau}, E.~T., {Kravtsov}, A.~V., \& {Nagai}, D. 2009, \apj, 705, 1129

\bibitem[{{Lelli} {et~al.}(2016{\natexlab{a}}){Lelli}, {McGaugh}, \&
  {Schombert}}]{SPARC}
{Lelli}, F., {McGaugh}, S.~S., \& {Schombert}, J.~M. 2016{\natexlab{a}}, \aj,
  152, 157

\bibitem[{{Lelli} {et~al.}(2016{\natexlab{b}}){Lelli}, {McGaugh}, \&
  {Schombert}}]{Lelli2016}
{Lelli}, F., {McGaugh}, S.~S., \& {Schombert}, J.~M. 2016{\natexlab{b}}, \apjl,
  816, L14

\bibitem[{{Lelli} {et~al.}(2017{\natexlab{a}}){Lelli}, {McGaugh}, \&
  {Schombert}}]{Lelli2017}
{Lelli}, F., {McGaugh}, S.~S., \& {Schombert}, J.~M. 2017{\natexlab{a}},
  \mnras, 468, L68

\bibitem[{{Lelli} {et~al.}(2017{\natexlab{b}}){Lelli}, {McGaugh}, {Schombert},
  \& {Pawlowski}}]{OneLaw}
{Lelli}, F., {McGaugh}, S.~S., {Schombert}, J.~M., \& {Pawlowski}, M.~S.
  2017{\natexlab{b}}, \apj, 836, 152

\bibitem[{{Li} {et~al.}(2018){Li}, {Lelli}, {McGaugh}, \& {Schombert}}]{Li2018}
{Li}, P., {Lelli}, F., {McGaugh}, S., \& {Schombert}, J. 2018, \aap, 615, A3

\bibitem[{{Limber} \& {Mathews}(1960)}]{Limber1960}
{Limber}, D.~N. \& {Mathews}, W.~G. 1960, \apj, 132, 286

\bibitem[{{Liu} {et~al.}(2023){Liu}, {Bulbul}, {Ramos-Ceja}, {Sanders},
  {Ghirardini}, {Bahar}, {Yeung}, {Gatuzz}, {Freyberg}, {Garrel}, {Zhang},
  {Merloni}, \& {Nandra}}]{Liu2023}
{Liu}, A., {Bulbul}, E., {Ramos-Ceja}, M.~E., {et~al.} 2023, \aap, 670, A96

\bibitem[{{Mamon} \& {{\L}okas}(2005)}]{Mamon2005II}
{Mamon}, G.~A. \& {{\L}okas}, E.~L. 2005, \mnras, 363, 705

\bibitem[{{Mathiesen} \& {Evrard}(2001)}]{Mathiesen2001}
{Mathiesen}, B.~F. \& {Evrard}, A.~E. 2001, \apj, 546, 100

\bibitem[{{McGaugh}(2005)}]{McGaugh2005}
{McGaugh}, S.~S. 2005, \apj, 632, 859

\bibitem[{{McGaugh} {et~al.}(2016){McGaugh}, {Lelli}, \&
  {Schombert}}]{McGaugh2016PRL}
{McGaugh}, S.~S., {Lelli}, F., \& {Schombert}, J.~M. 2016, Physical Review
  Letters, 117, 201101

\bibitem[{{McGaugh} {et~al.}(2021){McGaugh}, {Lelli}, {Schombert}, {Li},
  {Visgaitis}, {Parker}, \& {Pawlowski}}]{McGaugh2021}
{McGaugh}, S.~S., {Lelli}, F., {Schombert}, J.~M., {et~al.} 2021, \aj, 162, 202

\bibitem[{{Merrifield} \& {Kent}(1990)}]{Merrifield1990}
{Merrifield}, M.~R. \& {Kent}, S.~M. 1990, \aj, 99, 1548

\bibitem[{{Milgrom}(1983)}]{Milgrom1983}
{Milgrom}, M. 1983, \apj, 270, 365

\bibitem[{{Milgrom}(2015)}]{Milgrom2015}
{Milgrom}, M. 2015, \mnras, 454, 3810

\bibitem[{{Nagai} {et~al.}(2007){Nagai}, {Vikhlinin}, \&
  {Kravtsov}}]{Nagai2007}
{Nagai}, D., {Vikhlinin}, A., \& {Kravtsov}, A.~V. 2007, \apj, 655, 98

\bibitem[{{Navarro} {et~al.}(1996){Navarro}, {Eke}, \& {Frenk}}]{Navarro1996}
{Navarro}, J.~F., {Eke}, V.~R., \& {Frenk}, C.~S. 1996, \mnras, 283, L72

\bibitem[{{Nelson} {et~al.}(2014){Nelson}, {Lau}, \& {Nagai}}]{Nelson2014}
{Nelson}, K., {Lau}, E.~T., \& {Nagai}, D. 2014, \apj, 792, 25

\bibitem[{{Owers} {et~al.}(2011){Owers}, {Nulsen}, \& {Couch}}]{Owers2011}
{Owers}, M.~S., {Nulsen}, P. E.~J., \& {Couch}, W.~J. 2011, \apj, 741, 122

\bibitem[{{Planck Collaboration} {et~al.}(2014){Planck Collaboration}, {Ade},
  {Aghanim}, {Armitage-Caplan}, {Arnaud}, {Ashdown}, {Atrio-Barandela},
  {Aumont}, {Aussel}, {Baccigalupi}, {Banday}, {Barreiro}, {Barrena},
  {Bartelmann}, {Bartlett}, {Battaner}, {Benabed}, {Beno{\^\i}t},
  {Benoit-L{\'e}vy}, {Bernard}, {Bersanelli}, {Bielewicz}, {Bikmaev}, {Bobin},
  {Bock}, {B{\"o}hringer}, {Bonaldi}, {Bond}, {Borrill}, {Bouchet}, {Bridges},
  {Bucher}, {Burenin}, {Burigana}, {Butler}, {Cardoso}, {Carvalho}, {Catalano},
  {Challinor}, {Chamballu}, {Chary}, {Chen}, {Chiang}, {Chiang}, {Chon},
  {Christensen}, {Churazov}, {Church}, {Clements}, {Colombi}, {Colombo},
  {Comis}, {Couchot}, {Coulais}, {Crill}, {Curto}, {Cuttaia}, {Da Silva},
  {Dahle}, {Danese}, {Davies}, {Davis}, {de Bernardis}, {de Rosa}, {de Zotti},
  {Delabrouille}, {Delouis}, {D{\'e}mocl{\`e}s}, {D{\'e}sert}, {Dickinson},
  {Diego}, {Dolag}, {Dole}, {Donzelli}, {Dor{\'e}}, {Douspis}, {Dupac},
  {Efstathiou}, {Eisenhardt}, {En{\ss}lin}, {Eriksen}, {Feroz}, {Finelli},
  {Flores-Cacho}, {Forni}, {Frailis}, {Franceschi}, {Fromenteau}, {Galeotta},
  {Ganga}, {G{\'e}nova-Santos}, {Giard}, {Giardino}, {Gilfanov},
  {Giraud-H{\'e}raud}, {Gonz{\'a}lez-Nuevo}, {G{\'o}rski}, {Grainge},
  {Gratton}, {Gregorio}, {Groeneboom}, {Gruppuso}, {Hansen}, {Hanson},
  {Harrison}, {Hempel}, {Henrot-Versill{\'e}}, {Hern{\'a}ndez-Monteagudo},
  {Herranz}, {Hildebrandt}, {Hivon}, {Hobson}, {Holmes}, {Hornstrup}, {Hovest},
  {Huffenberger}, {Hurier}, {Hurley-Walker}, {Jaffe}, {Jaffe}, {Jones},
  {Juvela}, {Keih{\"a}nen}, {Keskitalo}, {Khamitov}, {Kisner}, {Kneissl},
  {Knoche}, {Knox}, {Kunz}, {Kurki-Suonio}, {Lagache}, {L{\"a}hteenm{\"a}ki},
  {Lamarre}, {Lasenby}, {Laureijs}, {Lawrence}, {Leahy}, {Leonardi},
  {Le{\'o}n-Tavares}, {Lesgourgues}, {Li}, {Liddle}, {Liguori}, {Lilje},
  {Linden-V{\o}rnle}, {L{\'o}pez-Caniego}, {Lubin}, {Mac{\'\i}as-P{\'e}rez},
  {MacTavish}, {Maffei}, {Maino}, {Mandolesi}, {Maris}, {Marshall}, {Martin},
  {Mart{\'\i}nez-Gonz{\'a}lez}, {Masi}, {Massardi}, {Matarrese}, {Matthai},
  {Mazzotta}, {Mei}, {Meinhold}, {Melchiorri}, {Melin}, {Mendes}, {Mennella},
  {Migliaccio}, {Mikkelsen}, {Mitra}, {Miville-Desch{\^e}nes}, {Moneti},
  {Montier}, {Morgante}, {Mortlock}, {Munshi}, {Murphy}, {Naselsky}, {Nati},
  {Natoli}, {Nesvadba}, {Netterfield}, {N{\o}rgaard-Nielsen}, {Noviello},
  {Novikov}, {Novikov}, {O'Dwyer}, {Olamaie}, {Osborne}, {Oxborrow}, {Paci},
  {Pagano}, {Pajot}, {Paoletti}, {Pasian}, {Patanchon}, {Pearson}, {Perdereau},
  {Perotto}, {Perrott}, {Perrotta}, {Piacentini}, {Piat}, {Pierpaoli},
  {Pietrobon}, {Plaszczynski}, {Pointecouteau}, {Polenta}, {Ponthieu}, {Popa},
  {Poutanen}, {Pratt}, {Pr{\'e}zeau}, {Prunet}, {Puget}, {Rachen}, {Reach},
  {Rebolo}, {Reinecke}, {Remazeilles}, {Renault}, {Ricciardi}, {Riller},
  {Ristorcelli}, {Rocha}, {Rosset}, {Roudier}, {Rowan-Robinson},
  {Rubi{\~n}o-Mart{\'\i}n}, {Rumsey}, {Rusholme}, {Sandri}, {Santos},
  {Saunders}, {Savini}, {Schammel}, {Scott}, {Seiffert}, {Shellard},
  {Shimwell}, {Spencer}, {Stanford}, {Starck}, {Stolyarov}, {Stompor},
  {Sudiwala}, {Sunyaev}, {Sureau}, {Sutton}, {Suur-Uski}, {Sygnet}, {Tauber},
  {Tavagnacco}, {Terenzi}, {Toffolatti}, {Tomasi}, {Tristram}, {Tucci},
  {Tuovinen}, {T{\"u}rler}, {Umana}, {Valenziano}, {Valiviita}, {Van Tent},
  {Vibert}, {Vielva}, {Villa}, {Vittorio}, {Wade}, {Wandelt}, {White}, {White},
  {Yvon}, {Zacchei}, \& {Zonca}}]{Planck2014XXIX}
{Planck Collaboration}, {Ade}, P.~A.~R., {Aghanim}, N., {et~al.} 2014, \aap,
  571, A29

\bibitem[{{Planck Collaboration} {et~al.}(2016{\natexlab{a}}){Planck
  Collaboration}, {Ade}, {Aghanim}, {Arnaud}, {Ashdown}, {Aumont},
  {Baccigalupi}, {Banday}, {Barreiro}, {Bartlett}, {Bartolo}, {Battaner},
  {Battye}, {Benabed}, {Beno{\^\i}t}, {Benoit-L{\'e}vy}, {Bernard},
  {Bersanelli}, {Bielewicz}, {Bock}, {Bonaldi}, {Bonavera}, {Bond}, {Borrill},
  {Bouchet}, {Boulanger}, {Bucher}, {Burigana}, {Butler}, {Calabrese},
  {Cardoso}, {Catalano}, {Challinor}, {Chamballu}, {Chary}, {Chiang}, {Chluba},
  {Christensen}, {Church}, {Clements}, {Colombi}, {Colombo}, {Combet},
  {Coulais}, {Crill}, {Curto}, {Cuttaia}, {Danese}, {Davies}, {Davis}, {de
  Bernardis}, {de Rosa}, {de Zotti}, {Delabrouille}, {D{\'e}sert}, {Di
  Valentino}, {Dickinson}, {Diego}, {Dolag}, {Dole}, {Donzelli}, {Dor{\'e}},
  {Douspis}, {Ducout}, {Dunkley}, {Dupac}, {Efstathiou}, {Elsner},
  {En{\ss}lin}, {Eriksen}, {Farhang}, {Fergusson}, {Finelli}, {Forni},
  {Frailis}, {Fraisse}, {Franceschi}, {Frejsel}, {Galeotta}, {Galli}, {Ganga},
  {Gauthier}, {Gerbino}, {Ghosh}, {Giard}, {Giraud-H{\'e}raud}, {Giusarma},
  {Gjerl{\o}w}, {Gonz{\'a}lez-Nuevo}, {G{\'o}rski}, {Gratton}, {Gregorio},
  {Gruppuso}, {Gudmundsson}, {Hamann}, {Hansen}, {Hanson}, {Harrison}, {Helou},
  {Henrot-Versill{\'e}}, {Hern{\'a}ndez-Monteagudo}, {Herranz}, {Hildebrandt},
  {Hivon}, {Hobson}, {Holmes}, {Hornstrup}, {Hovest}, {Huang}, {Huffenberger},
  {Hurier}, {Jaffe}, {Jaffe}, {Jones}, {Juvela}, {Keih{\"a}nen}, {Keskitalo},
  {Kisner}, {Kneissl}, {Knoche}, {Knox}, {Kunz}, {Kurki-Suonio}, {Lagache},
  {L{\"a}hteenm{\"a}ki}, {Lamarre}, {Lasenby}, {Lattanzi}, {Lawrence}, {Leahy},
  {Leonardi}, {Lesgourgues}, {Levrier}, {Lewis}, {Liguori}, {Lilje},
  {Linden-V{\o}rnle}, {L{\'o}pez-Caniego}, {Lubin}, {Mac{\'\i}as-P{\'e}rez},
  {Maggio}, {Maino}, {Mandolesi}, {Mangilli}, {Marchini}, {Maris}, {Martin},
  {Martinelli}, {Mart{\'\i}nez-Gonz{\'a}lez}, {Masi}, {Matarrese}, {McGehee},
  {Meinhold}, {Melchiorri}, {Melin}, {Mendes}, {Mennella}, {Migliaccio},
  {Millea}, {Mitra}, {Miville-Desch{\^e}nes}, {Moneti}, {Montier}, {Morgante},
  {Mortlock}, {Moss}, {Munshi}, {Murphy}, {Naselsky}, {Nati}, {Natoli},
  {Netterfield}, {N{\o}rgaard-Nielsen}, {Noviello}, {Novikov}, {Novikov},
  {Oxborrow}, {Paci}, {Pagano}, {Pajot}, {Paladini}, {Paoletti}, {Partridge},
  {Pasian}, {Patanchon}, {Pearson}, {Perdereau}, {Perotto}, {Perrotta},
  {Pettorino}, {Piacentini}, {Piat}, {Pierpaoli}, {Pietrobon}, {Plaszczynski},
  {Pointecouteau}, {Polenta}, {Popa}, {Pratt}, {Pr{\'e}zeau}, {Prunet},
  {Puget}, {Rachen}, {Reach}, {Rebolo}, {Reinecke}, {Remazeilles}, {Renault},
  {Renzi}, {Ristorcelli}, {Rocha}, {Rosset}, {Rossetti}, {Roudier},
  {Rouill{\'e} d'Orfeuil}, {Rowan-Robinson}, {Rubi{\~n}o-Mart{\'\i}n},
  {Rusholme}, {Said}, {Salvatelli}, {Salvati}, {Sandri}, {Santos},
  {Savelainen}, {Savini}, {Scott}, {Seiffert}, {Serra}, {Shellard}, {Spencer},
  {Spinelli}, {Stolyarov}, {Stompor}, {Sudiwala}, {Sunyaev}, {Sutton},
  {Suur-Uski}, {Sygnet}, {Tauber}, {Terenzi}, {Toffolatti}, {Tomasi},
  {Tristram}, {Trombetti}, {Tucci}, {Tuovinen}, {T{\"u}rler}, {Umana},
  {Valenziano}, {Valiviita}, {Van Tent}, {Vielva}, {Villa}, {Wade}, {Wandelt},
  {Wehus}, {White}, {White}, {Wilkinson}, {Yvon}, {Zacchei}, \&
  {Zonca}}]{Planck2016XIII}
{Planck Collaboration}, {Ade}, P.~A.~R., {Aghanim}, N., {et~al.}
  2016{\natexlab{a}}, \aap, 594, A13

\bibitem[{{Planck Collaboration} {et~al.}(2016{\natexlab{b}}){Planck
  Collaboration}, {Ade}, {Aghanim}, {Arnaud}, {Ashdown}, {Aumont},
  {Baccigalupi}, {Banday}, {Barreiro}, {Bartlett}, {Bartolo}, {Battaner},
  {Battye}, {Benabed}, {Beno{\^\i}t}, {Benoit-L{\'e}vy}, {Bernard},
  {Bersanelli}, {Bielewicz}, {Bock}, {Bonaldi}, {Bonavera}, {Bond}, {Borrill},
  {Bouchet}, {Bucher}, {Burigana}, {Butler}, {Calabrese}, {Cardoso},
  {Catalano}, {Challinor}, {Chamballu}, {Chary}, {Chiang}, {Christensen},
  {Church}, {Clements}, {Colombi}, {Colombo}, {Combet}, {Comis}, {Couchot},
  {Coulais}, {Crill}, {Curto}, {Cuttaia}, {Danese}, {Davies}, {Davis}, {de
  Bernardis}, {de Rosa}, {de Zotti}, {Delabrouille}, {D{\'e}sert}, {Diego},
  {Dolag}, {Dole}, {Donzelli}, {Dor{\'e}}, {Douspis}, {Ducout}, {Dupac},
  {Efstathiou}, {Elsner}, {En{\ss}lin}, {Eriksen}, {Falgarone}, {Fergusson},
  {Finelli}, {Forni}, {Frailis}, {Fraisse}, {Franceschi}, {Frejsel},
  {Galeotta}, {Galli}, {Ganga}, {Giard}, {Giraud-H{\'e}raud}, {Gjerl{\o}w},
  {Gonz{\'a}lez-Nuevo}, {G{\'o}rski}, {Gratton}, {Gregorio}, {Gruppuso},
  {Gudmundsson}, {Hansen}, {Hanson}, {Harrison}, {Henrot-Versill{\'e}},
  {Hern{\'a}ndez-Monteagudo}, {Herranz}, {Hildebrandt}, {Hivon}, {Hobson},
  {Holmes}, {Hornstrup}, {Hovest}, {Huffenberger}, {Hurier}, {Jaffe}, {Jaffe},
  {Jones}, {Juvela}, {Keih{\"a}nen}, {Keskitalo}, {Kisner}, {Kneissl},
  {Knoche}, {Kunz}, {Kurki-Suonio}, {Lagache}, {L{\"a}hteenm{\"a}ki},
  {Lamarre}, {Lasenby}, {Lattanzi}, {Lawrence}, {Leonardi}, {Lesgourgues},
  {Levrier}, {Liguori}, {Lilje}, {Linden-V{\o}rnle}, {L{\'o}pez-Caniego},
  {Lubin}, {Mac{\'\i}as-P{\'e}rez}, {Maggio}, {Maino}, {Mandolesi}, {Mangilli},
  {Maris}, {Martin}, {Mart{\'\i}nez-Gonz{\'a}lez}, {Masi}, {Matarrese},
  {McGehee}, {Meinhold}, {Melchiorri}, {Melin}, {Mendes}, {Mennella},
  {Migliaccio}, {Mitra}, {Miville-Desch{\^e}nes}, {Moneti}, {Montier},
  {Morgante}, {Mortlock}, {Moss}, {Munshi}, {Murphy}, {Naselsky}, {Nati},
  {Natoli}, {Netterfield}, {N{\o}rgaard-Nielsen}, {Noviello}, {Novikov},
  {Novikov}, {Oxborrow}, {Paci}, {Pagano}, {Pajot}, {Paoletti}, {Partridge},
  {Pasian}, {Patanchon}, {Pearson}, {Perdereau}, {Perotto}, {Perrotta},
  {Pettorino}, {Piacentini}, {Piat}, {Pierpaoli}, {Pietrobon}, {Plaszczynski},
  {Pointecouteau}, {Polenta}, {Popa}, {Pratt}, {Pr{\'e}zeau}, {Prunet},
  {Puget}, {Rachen}, {Rebolo}, {Reinecke}, {Remazeilles}, {Renault}, {Renzi},
  {Ristorcelli}, {Rocha}, {Roman}, {Rosset}, {Rossetti}, {Roudier},
  {Rubi{\~n}o-Mart{\'\i}n}, {Rusholme}, {Sandri}, {Santos}, {Savelainen},
  {Savini}, {Scott}, {Seiffert}, {Shellard}, {Spencer}, {Stolyarov}, {Stompor},
  {Sudiwala}, {Sunyaev}, {Sutton}, {Suur-Uski}, {Sygnet}, {Tauber}, {Terenzi},
  {Toffolatti}, {Tomasi}, {Tristram}, {Tucci}, {Tuovinen}, {T{\"u}rler},
  {Umana}, {Valenziano}, {Valiviita}, {Van Tent}, {Vielva}, {Villa}, {Wade},
  {Wandelt}, {Wehus}, {Weller}, {White}, {Yvon}, {Zacchei}, \&
  {Zonca}}]{Planck2016XXIV}
{Planck Collaboration}, {Ade}, P.~A.~R., {Aghanim}, N., {et~al.}
  2016{\natexlab{b}}, \aap, 594, A24

\bibitem[{{Plummer}(1911)}]{Plummer1911}
{Plummer}, H.~C. 1911, \mnras, 71, 460

\bibitem[{{Pontzen} {et~al.}(2015){Pontzen}, {Read}, {Teyssier}, {Governato},
  {Gualandris}, {Roth}, \& {Devriendt}}]{Pontzen2015}
{Pontzen}, A., {Read}, J.~I., {Teyssier}, R., {et~al.} 2015, \mnras, 451, 1366

\bibitem[{{Postman} {et~al.}(2012){Postman}, {Coe}, {Ben{\'\i}tez}, {Bradley},
  {Broadhurst}, {Donahue}, {Ford}, {Graur}, {Graves}, {Jouvel}, {Koekemoer},
  {Lemze}, {Medezinski}, {Molino}, {Moustakas}, {Ogaz}, {Riess}, {Rodney},
  {Rosati}, {Umetsu}, {Zheng}, {Zitrin}, {Bartelmann}, {Bouwens}, {Czakon},
  {Golwala}, {Host}, {Infante}, {Jha}, {Jimenez-Teja}, {Kelson}, {Lahav},
  {Lazkoz}, {Maoz}, {McCully}, {Melchior}, {Meneghetti}, {Merten}, {Moustakas},
  {Nonino}, {Patel}, {Reg{\"o}s}, {Sayers}, {Seitz}, \& {Van der
  Wel}}]{Postman2012}
{Postman}, M., {Coe}, D., {Ben{\'\i}tez}, N., {et~al.} 2012, \apjs, 199, 25

\bibitem[{{Read} {et~al.}(2016){Read}, {Agertz}, \& {Collins}}]{Read2016}
{Read}, J.~I., {Agertz}, O., \& {Collins}, M.~L.~M. 2016, \mnras, 459, 2573

\bibitem[{{Read} {et~al.}(2021){Read}, {Mamon}, {Vasiliev}, {Watkins},
  {Walker}, {Pe{\~n}arrubia}, {Wilkinson}, {Dehnen}, \& {Das}}]{Read2021}
{Read}, J.~I., {Mamon}, G.~A., {Vasiliev}, E., {et~al.} 2021, \mnras, 501, 978

\bibitem[{{Read} \& {Steger}(2017)}]{Read2017}
{Read}, J.~I. \& {Steger}, P. 2017, \mnras, 471, 4541

\bibitem[{{Read} {et~al.}(2018){Read}, {Walker}, \& {Steger}}]{Read2018}
{Read}, J.~I., {Walker}, M.~G., \& {Steger}, P. 2018, \mnras, 481, 860

\bibitem[{{Read} {et~al.}(2006){Read}, {Wilkinson}, {Evans}, {Gilmore}, \&
  {Kleyna}}]{Read2006}
{Read}, J.~I., {Wilkinson}, M.~I., {Evans}, N.~W., {Gilmore}, G., \& {Kleyna},
  J.~T. 2006, \mnras, 367, 387

\bibitem[{{Reiprich} \& {B{\"o}hringer}(2002)}]{Reiprich2002}
{Reiprich}, T.~H. \& {B{\"o}hringer}, H. 2002, \apj, 567, 716

\bibitem[{{Sanders}(2003)}]{Sanders2003}
{Sanders}, R.~H. 2003, \mnras, 342, 901

\bibitem[{{Schombert} {et~al.}(2022){Schombert}, {McGaugh}, \&
  {Lelli}}]{Schombert2022}
{Schombert}, J., {McGaugh}, S., \& {Lelli}, F. 2022, \aj, 163, 154

\bibitem[{{Serra} {et~al.}(2011){Serra}, {Diaferio}, {Murante}, \&
  {Borgani}}]{Serra2011}
{Serra}, A.~L., {Diaferio}, A., {Murante}, G., \& {Borgani}, S. 2011, \mnras,
  412, 800

\bibitem[{{Shelest} \& {Lelli}(2020)}]{Shelest2020}
{Shelest}, A. \& {Lelli}, F. 2020, \aap, 641, A31

\bibitem[{{Skrutskie} {et~al.}(2006){Skrutskie}, {Cutri}, {Stiening},
  {Weinberg}, {Schneider}, {Carpenter}, {Beichman}, {Capps}, {Chester},
  {Elias}, {Huchra}, {Liebert}, {Lonsdale}, {Monet}, {Price}, {Seitzer},
  {Jarrett}, {Kirkpatrick}, {Gizis}, {Howard}, {Evans}, {Fowler}, {Fullmer},
  {Hurt}, {Light}, {Kopan}, {Marsh}, {McCallon}, {Tam}, {Van Dyk}, \&
  {Wheelock}}]{Skrutskie2006}
{Skrutskie}, M.~F., {Cutri}, R.~M., {Stiening}, R., {et~al.} 2006, \aj, 131,
  1163

\bibitem[{{Sohn} {et~al.}(2017){Sohn}, {Geller}, {Zahid}, {Fabricant},
  {Diaferio}, \& {Rines}}]{Sohn2017}
{Sohn}, J., {Geller}, M.~J., {Zahid}, H.~J., {et~al.} 2017, \apjs, 229, 20

\bibitem[{{Tian} {et~al.}(2020){Tian}, {Umetsu}, {Ko}, {Donahue}, \&
  {Chiu}}]{Tian2020}
{Tian}, Y., {Umetsu}, K., {Ko}, C.-M., {Donahue}, M., \& {Chiu}, I.~N. 2020,
  \apj, 896, 70

\bibitem[{{Tian} {et~al.}(2021){Tian}, {Yu}, {Li}, {McGaugh}, \&
  {Ko}}]{Tian2021}
{Tian}, Y., {Yu}, P.-C., {Li}, P., {McGaugh}, S.~S., \& {Ko}, C.-M. 2021, \apj,
  910, 56

\bibitem[{{Tiret} {et~al.}(2007){Tiret}, {Combes}, {Angus}, {Famaey}, \&
  {Zhao}}]{Tiret2007}
{Tiret}, O., {Combes}, F., {Angus}, G.~W., {Famaey}, B., \& {Zhao}, H.~S. 2007,
  \aap, 476, L1

\bibitem[{{Umetsu} {et~al.}(2016){Umetsu}, {Zitrin}, {Gruen}, {Merten},
  {Donahue}, \& {Postman}}]{Umetsu2016}
{Umetsu}, K., {Zitrin}, A., {Gruen}, D., {et~al.} 2016, \apj, 821, 116

\bibitem[{{van der Marel}(1994)}]{vanderMarel1994}
{van der Marel}, R.~P. 1994, \mnras, 270, 271

\bibitem[{{Ventimiglia} {et~al.}(2008){Ventimiglia}, {Voit}, {Donahue}, \&
  {Ameglio}}]{Ventimiglia2008}
{Ventimiglia}, D.~A., {Voit}, G.~M., {Donahue}, M., \& {Ameglio}, S. 2008,
  \apj, 685, 118

\bibitem[{{Vikhlinin} {et~al.}(2009){Vikhlinin}, {Burenin}, {Ebeling},
  {Forman}, {Hornstrup}, {Jones}, {Kravtsov}, {Murray}, {Nagai}, {Quintana}, \&
  {Voevodkin}}]{Vikhlinin2009}
{Vikhlinin}, A., {Burenin}, R.~A., {Ebeling}, H., {et~al.} 2009, \apj, 692,
  1033

\bibitem[{{Vikhlinin} {et~al.}(2006){Vikhlinin}, {Kravtsov}, {Forman}, {Jones},
  {Markevitch}, {Murray}, \& {Van Speybroeck}}]{Vikhlinin2006}
{Vikhlinin}, A., {Kravtsov}, A., {Forman}, W., {et~al.} 2006, \apj, 640, 691

\bibitem[{{Wenger} {et~al.}(2000){Wenger}, {Ochsenbein}, {Egret}, {Dubois},
  {Bonnarel}, {Borde}, {Genova}, {Jasniewicz}, {Lalo{\"e}}, {Lesteven}, \&
  {Monier}}]{Wenger2000}
{Wenger}, M., {Ochsenbein}, F., {Egret}, D., {et~al.} 2000, \aaps, 143, 9

\bibitem[{{Zhang} {et~al.}(2011){Zhang}, {Andernach}, {Caretta}, {Reiprich},
  {B{\"o}hringer}, {Puchwein}, {Sijacki}, \& {Girardi}}]{Zhang2011}
{Zhang}, Y.~Y., {Andernach}, H., {Caretta}, C.~A., {et~al.} 2011, \aap, 526,
  A105

\bibitem[{{Zhao} \& {Famaey}(2012)}]{Zhao2012}
{Zhao}, H. \& {Famaey}, B. 2012, \prd, 86, 067301

\bibitem[{{Zwicky}(1933)}]{Zwicky1933}
{Zwicky}, F. 1933, Helvetica Physica Acta, 6, 110

\end{thebibliography}

\begin{appendix}
\FloatBarrier
\section{The mass profiles of A2029 assuming different, radially varying, completeness functions}\label{sec:incompleteness}

In order to quantitatively study the possible bias due to the radially varying completeness, we derive the mass profiles for A2029 assuming different completeness functions. For simplicity, we assume the completeness of A2029 linearly decreases from 100\% at the innermost radius to 60\% and 20\% at the outermost radius, which correspond to 40\% and 80\% radial variations in the completeness profiles, respectively. Since we cannot add missing galaxies at large radii, we can instead proceed by randomly rejecting some galaxies at small radii to achieve a constant completeness profile. However, rejecting a large number of galaxies could lower the statistical robustness for the velocity dispersion measurements. So we keep all these galaxies, but lower its membership probability from 100\% to
\begin{equation}
    P_{\rm membership} = (1-P) + P\times\frac{R-R_{\rm min}}{R_{\rm max}-R_{\rm min}},
\end{equation}
where $P=40\%$ for a 40\% variation of completeness, and $P=80\%$ for a 80\% variation; $R$ is the projected radius of each cluster galaxy; $R_{\rm max}$ and $R_{\rm min}$ refer to the radii of the innermost and outermost galaxies, respectively. The designed function of membership probability reduces the effective number of cluster galaxies while preserving all the galaxies for velocity dispersion measurements. 

\begin{figure}
    \centering
    \includegraphics[scale=0.45]{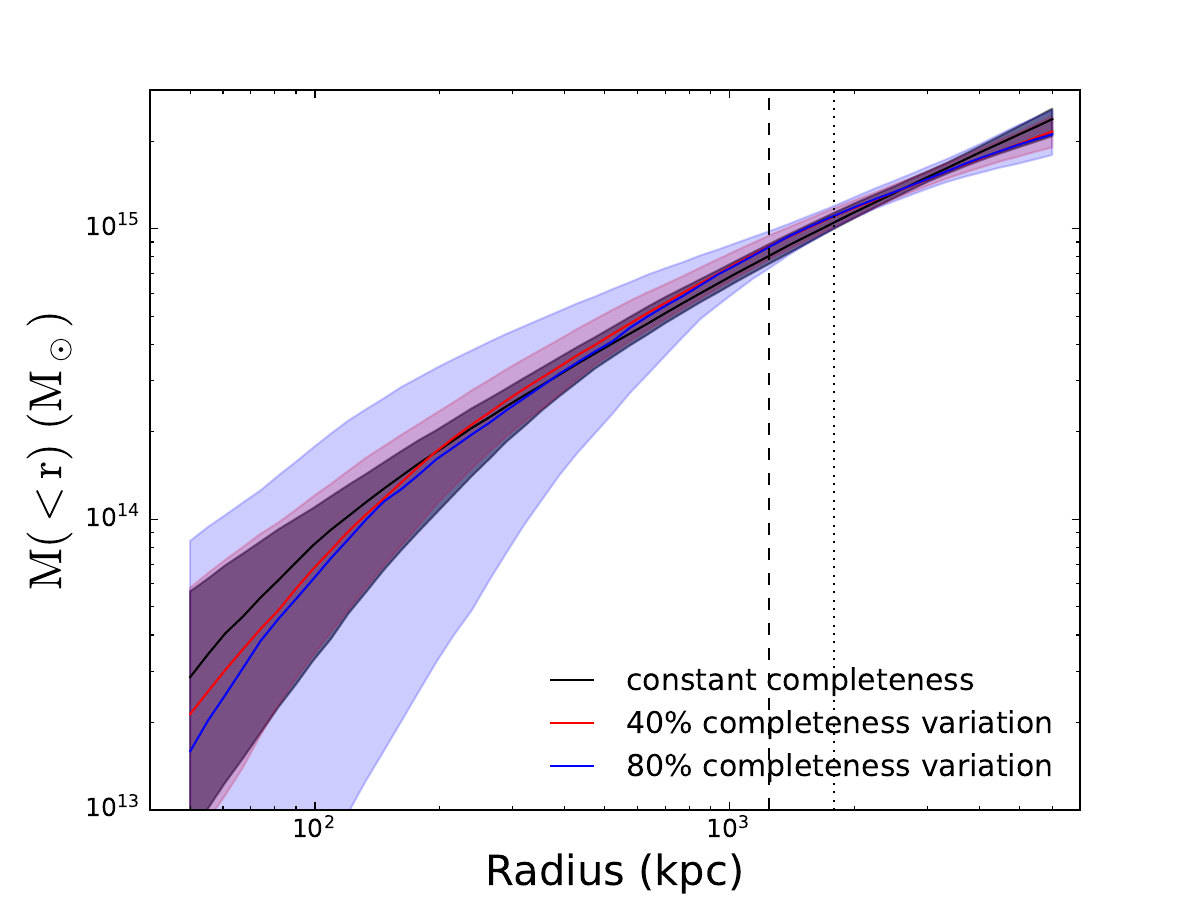}
    \caption{Mass profiles of A2029 with radially varying completeness. The black line assumes the completeness is a constant with radius. The red and blue lines assume the completeness decreases from the innermost region to the ourtermost region by 40\% and 80\%, respectively. Shadow regions present the 1 $\sigma$ credible intervals. Larger completeness variations show larger uncertainties due to the smaller sample after reducing the effective number of galaxies at small radii. The vertical dashed and dotted lines mark the positions of $r_{500}$ and $R_{\rm half}$, respectively. The variations of the mass profiles due to different completeness functions are well with the 1 $\sigma$ region, suggesting our results are not sensitive to incompleteness.}
    \label{fig:completeness}
\end{figure}

Figure \ref{fig:completeness} shows the mass profiles of A2029 when assuming different radial variations of completeness. At small radii, the enclosed mass becomes smaller for a larger radial variation of completeness. However, the uncertainties also increase quickly. This is because the galaxy sample becomes smaller after we reduce the effective number of galaxies by lowering the membership probability. The variation in the mass profile is smaller than the uncertainty. At large radii, the mass profiles are even more consistent. This suggests that incompleteness does not affect our mass measurements significantly. Our mass profiles are robust within their errors. 

\section{Example corner plots for the mass-velocity anisotropy degeneracy}\label{sec:corner}

\begin{figure*}
    \centering
    \includegraphics[scale=0.65]{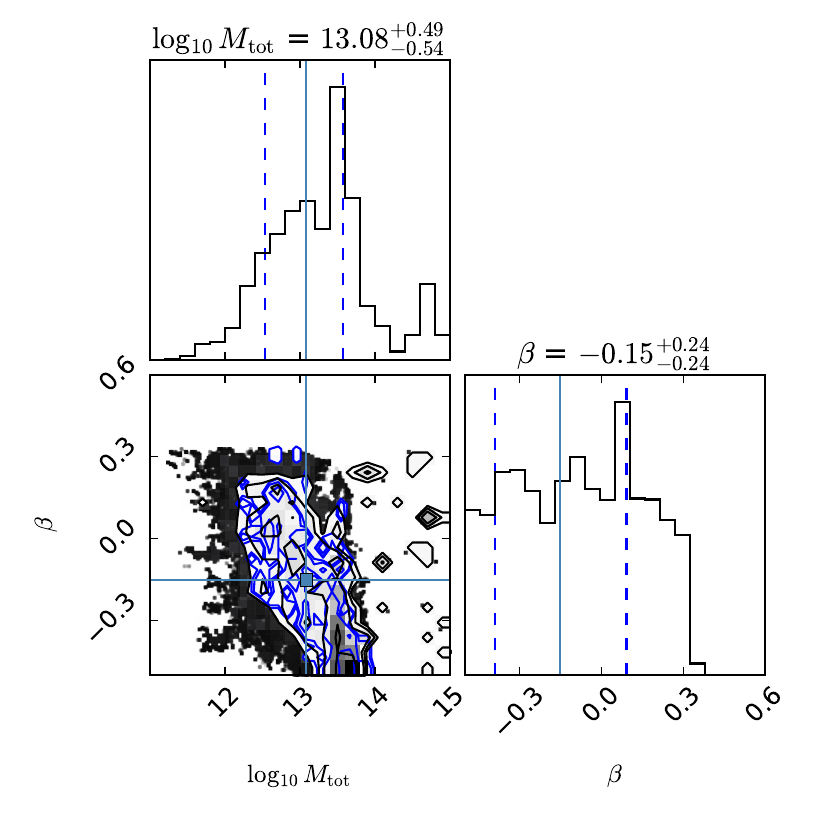}\includegraphics[scale=0.65]{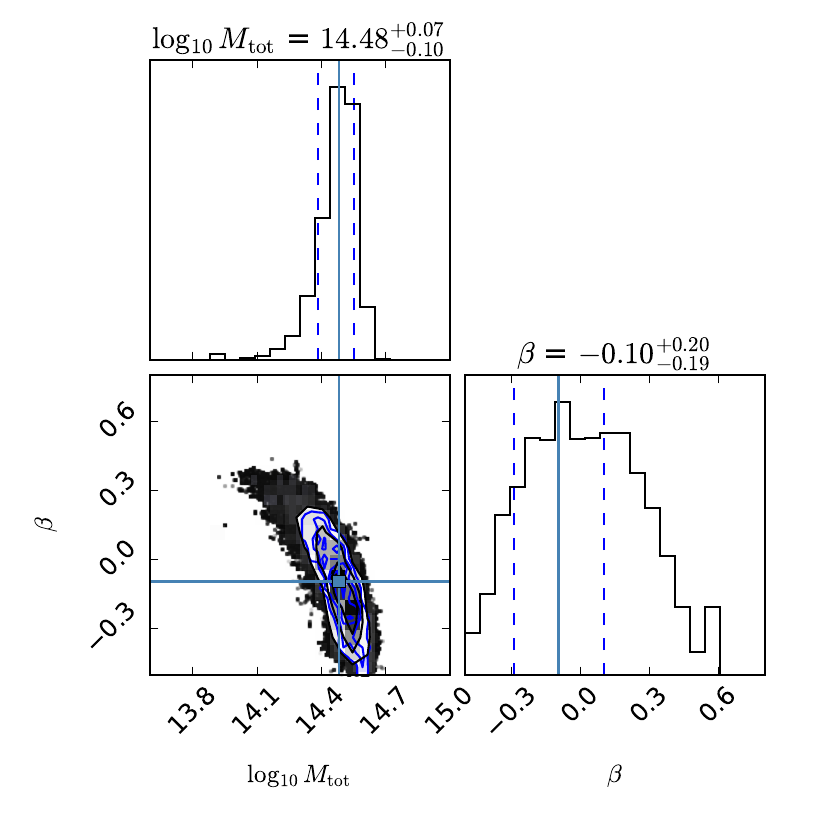}\\
    \includegraphics[scale=0.65]{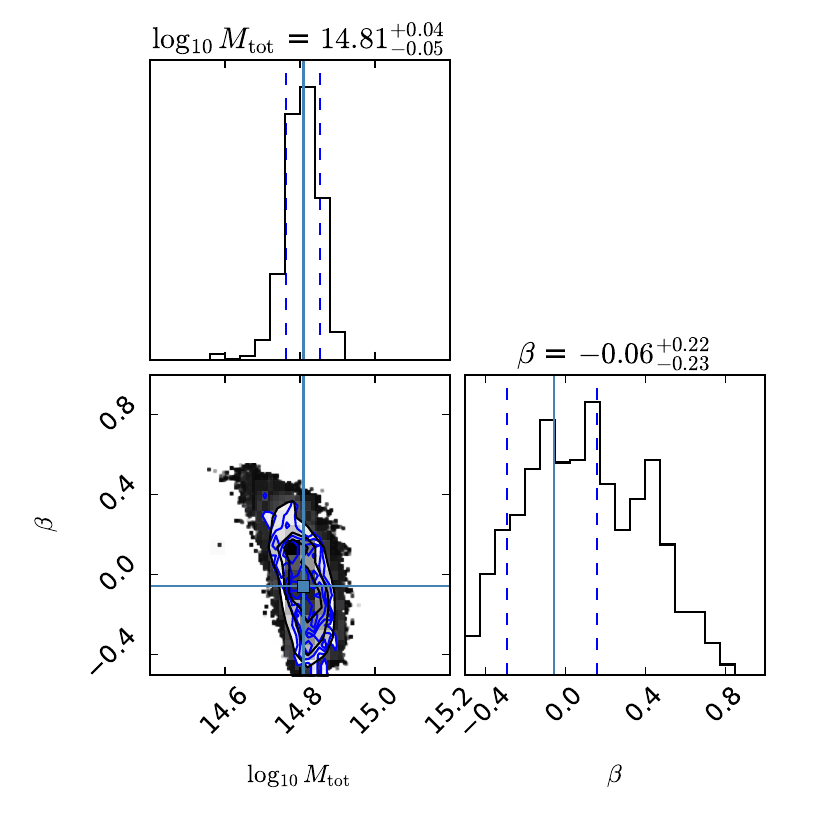}\includegraphics[scale=0.65]{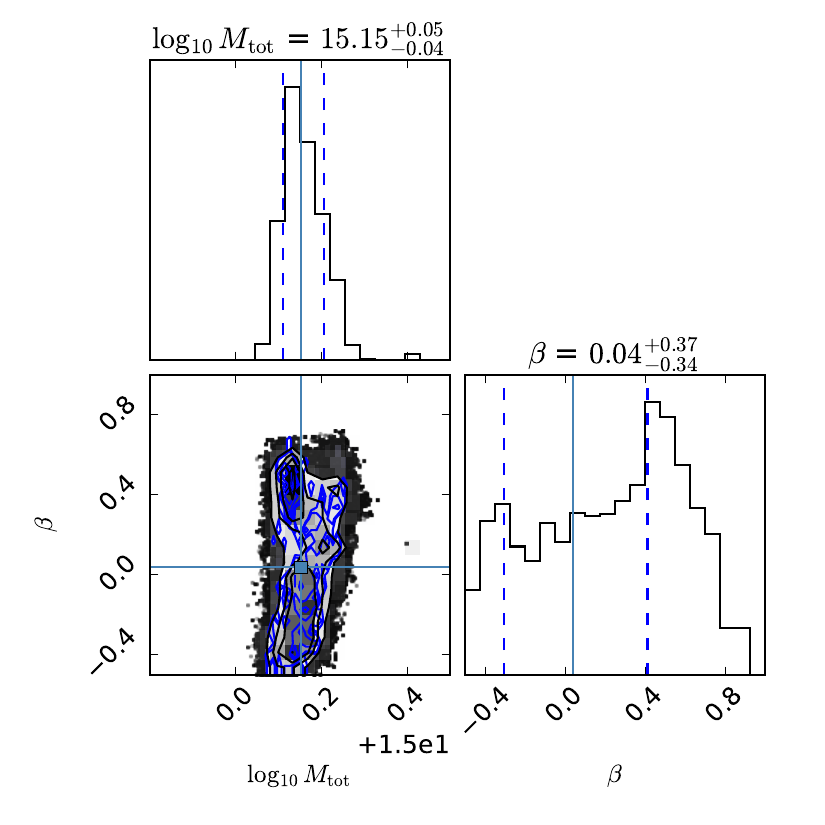}
    \caption{Example corner plots for cluster A0085 showing the mass-velocity anisotropy degeneracy. The four panels present the posterior distributions of total enclosed mass and velocity anisotropy at four different radii where binned velocity dispersion data are available: $r=96$ kpc (top left), $r=570$ kpc (top right), $r=965$ kpc (bottom left), and $r=1827$ kpc (bottom right). Blue color marks regions that are used for parameter estimations. Blue crosses indicate the position of the best-fit parameters and vertical dashed lines outline the 1 $\sigma$ regions. The total enclosed mass is well determined at all radii, while the velocity anisotropy is less constrained at small radii but gets slightly better towards large radii.}
    \label{fig:corner}
\end{figure*}

We examine how the two virial shape parameters help ameliorate the mass-velocity anisotropy degeneracy by plotting their posterior distributions in Figure \ref{fig:corner}. Since both the enclosed total mass and velocity anisotropy are functions of radius, we present four posterior distributions nearly equally spanning from the innermost to ourtermost radii for the example cluster A0085 (other clusters present similar posterior distributions). Figure \ref{fig:corner} shows that the enclosed total mass is well constrained at all radii, suggesting the two virial shape parameters indeed help ameliorate the degeneracy. In contrast, velocity anisotropy presents wide distributions. At large radii, we observe some Gaussian-like distributions, but they generally have a large standard deviation. This suggests that the degeneracy, though ameliorated by virial shape parameters, is not completely broken.

\section{Total density profiles of ten relaxed clusters.}
\FloatBarrier
\begin{figure*}
    \centering
    \includegraphics[scale=0.45]{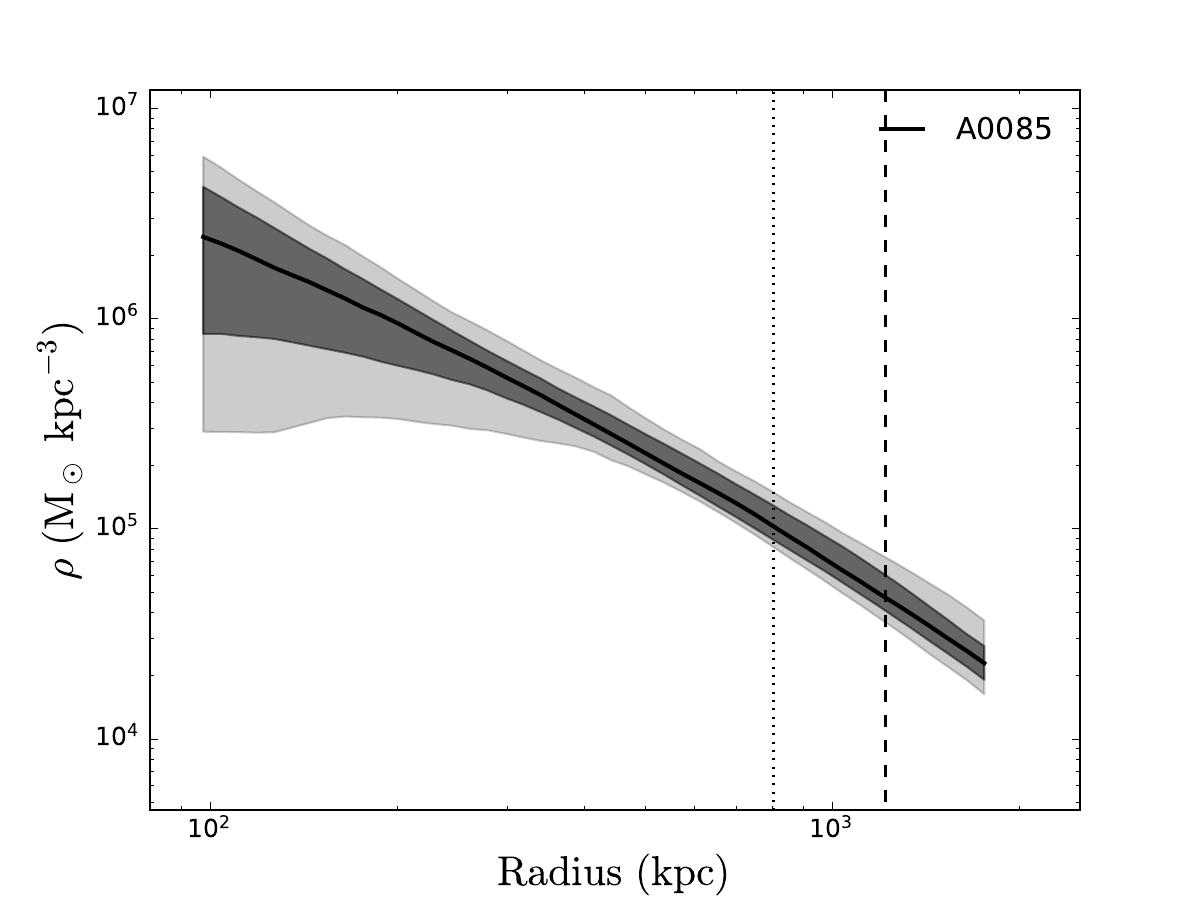}\includegraphics[scale=0.45]{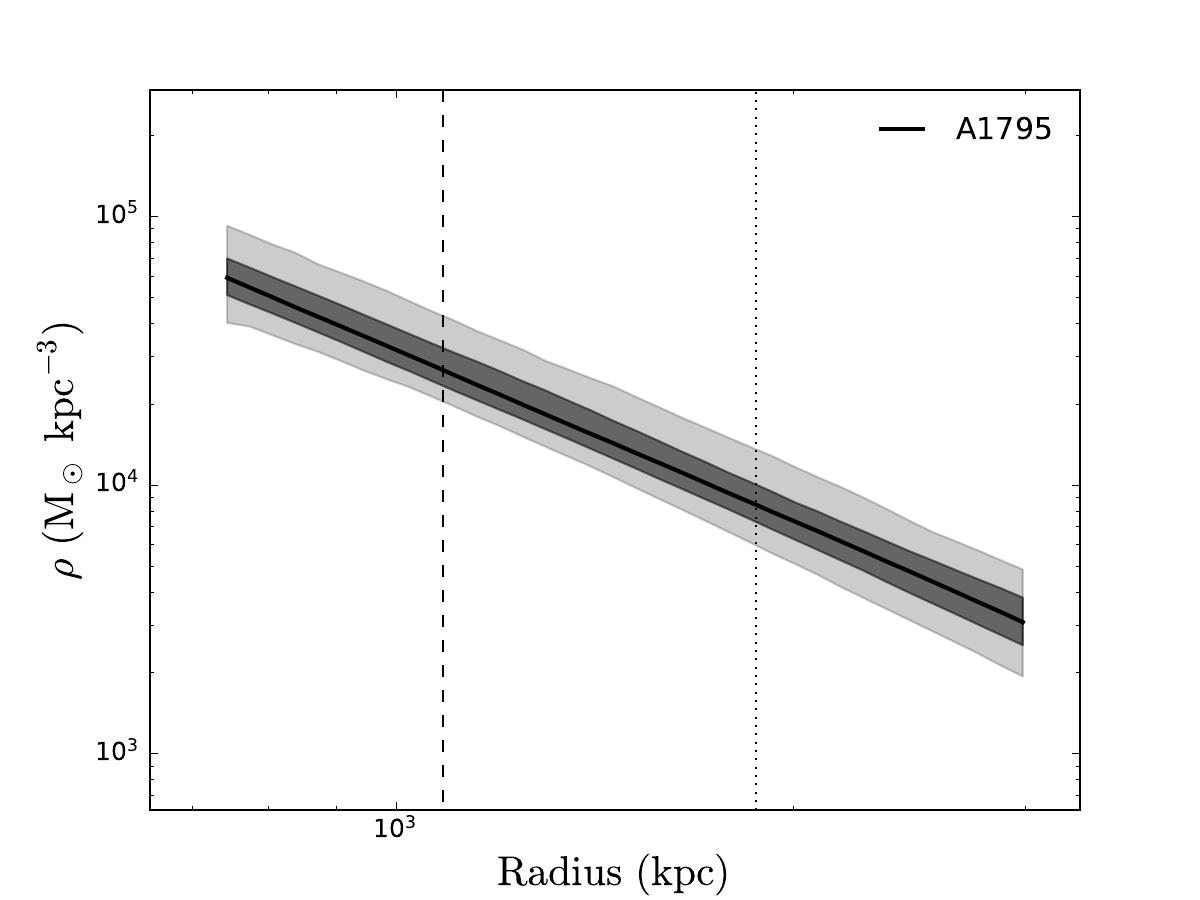}\\
    \includegraphics[scale=0.45]{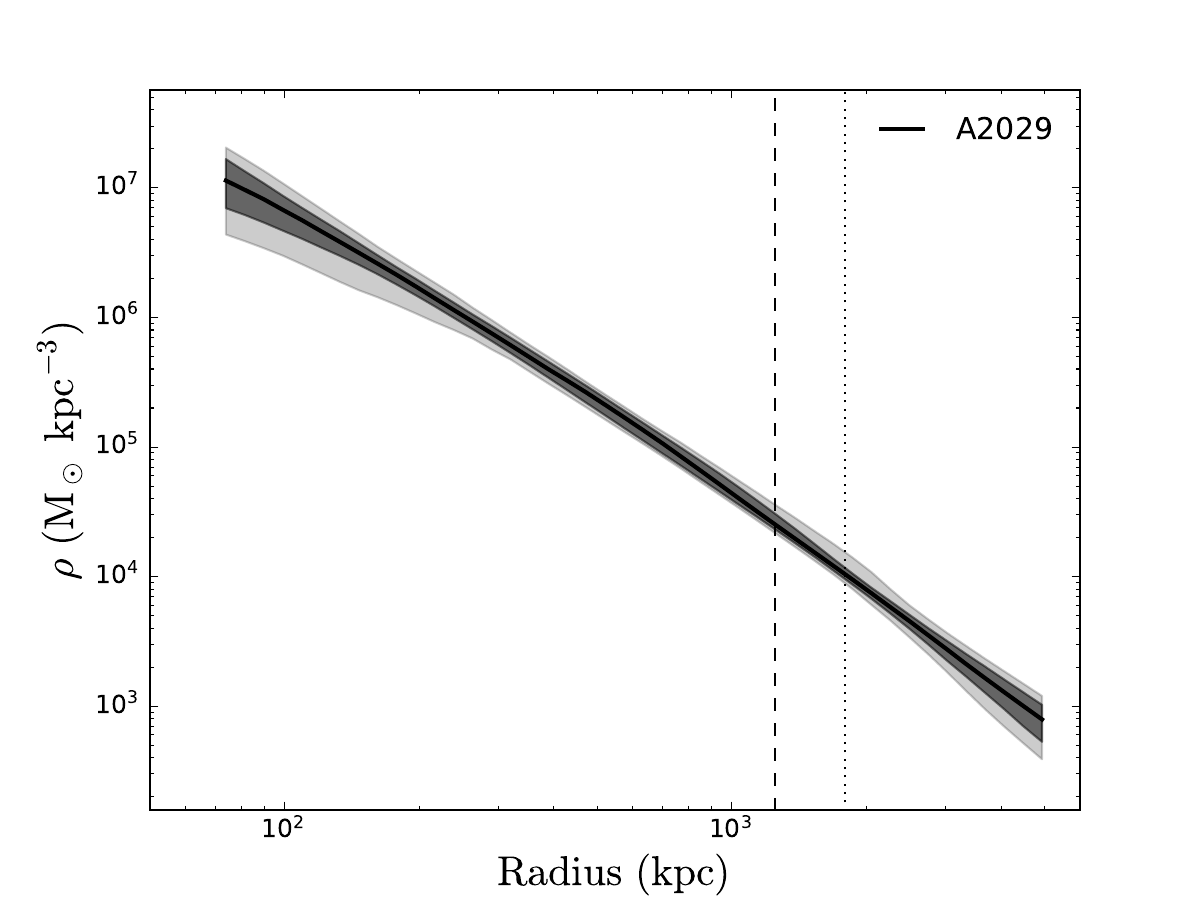}\includegraphics[scale=0.45]{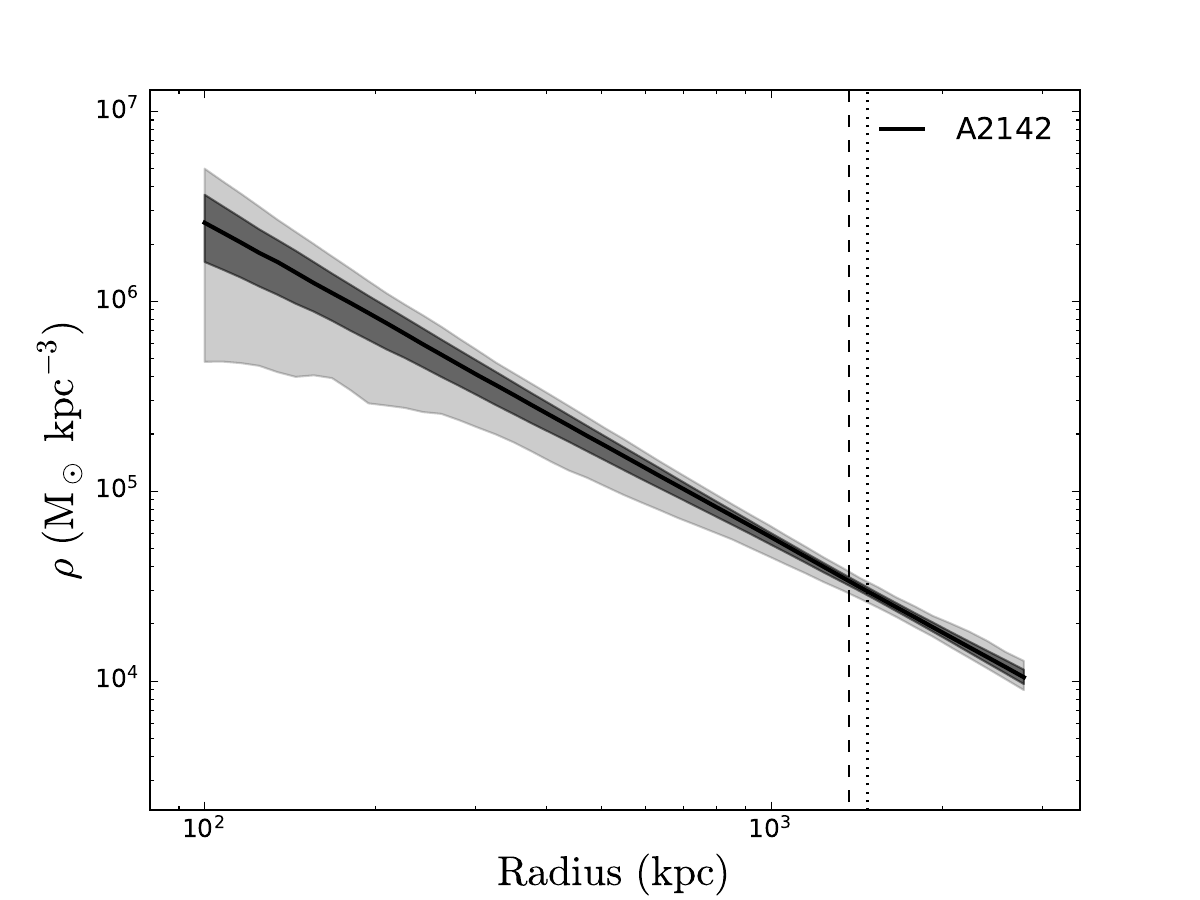}\\
    \includegraphics[scale=0.45]{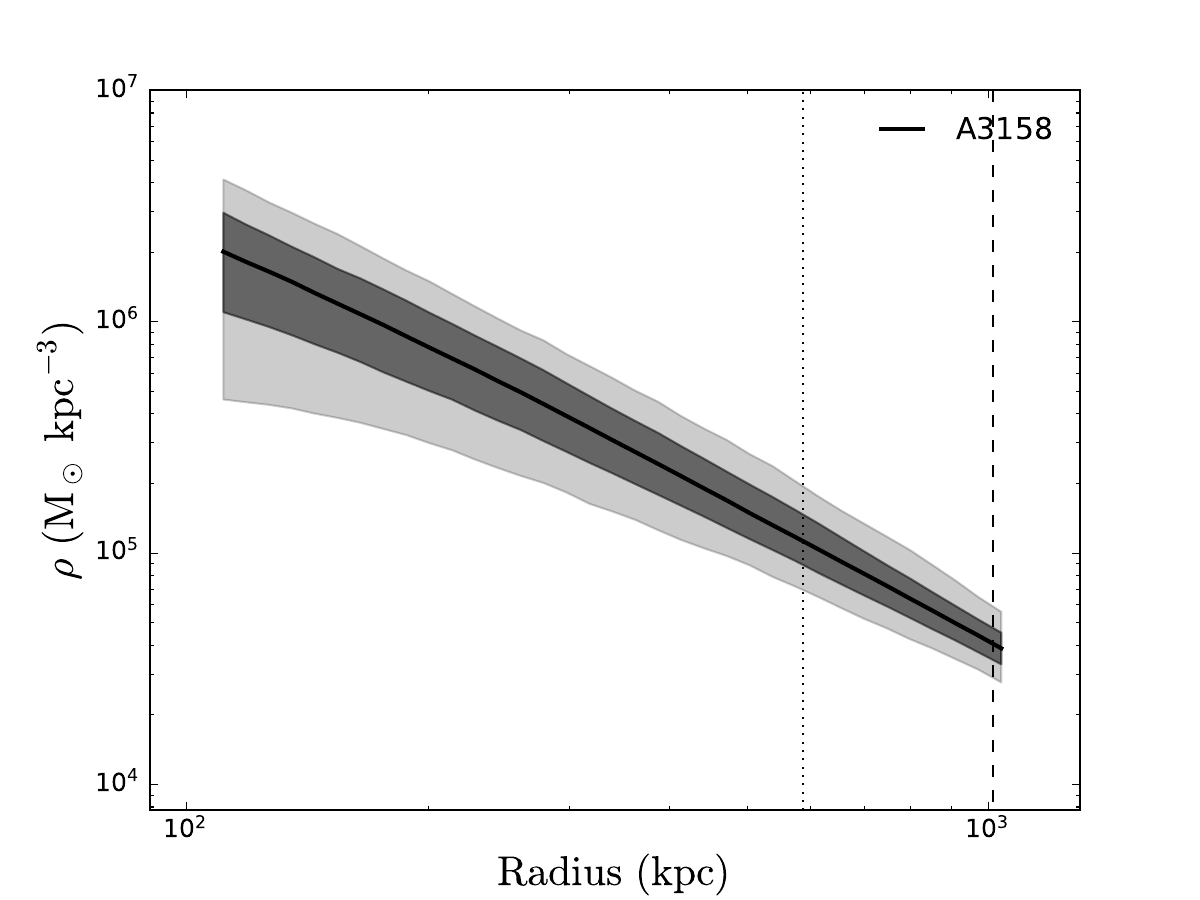}\includegraphics[scale=0.45]{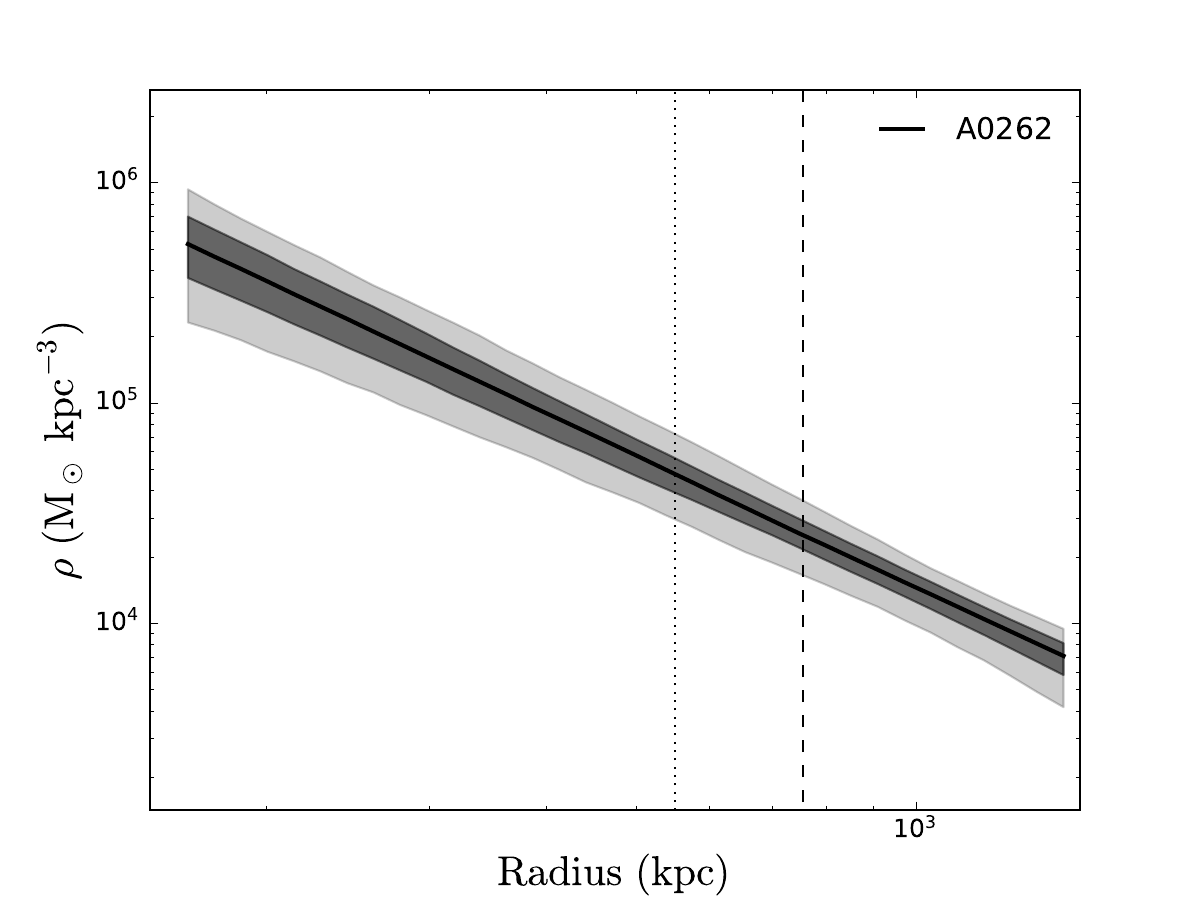}\\
    \caption{Density profiles corresponding to the total cumulative mass profiles in Figure \ref{fig:MassProfile}. Dark and light shadow area are 1$\sigma$ and 2$\sigma$ regions. Vertical dashed and dotted lines indicate the positions of $r_{500}$ from X-ray data and $R_{\rm half}$ from optical data. Note that these are not dark matter density profiles but the total density profiles including baryonic contributions. }
    \label{fig:DensityProfile}
\end{figure*}
\renewcommand{\thefigure}{C.1}
\begin{figure*}
    \centering
    \includegraphics[scale=0.45]{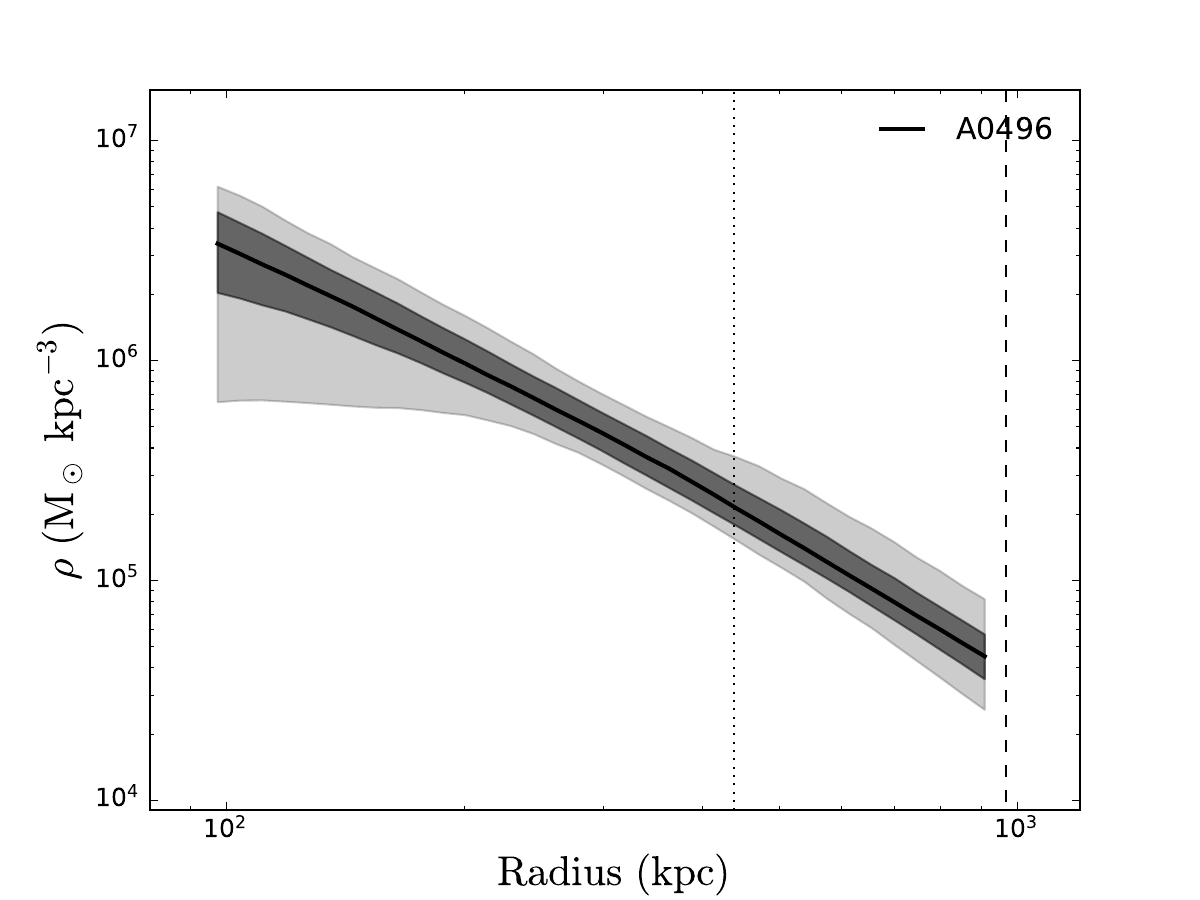}\includegraphics[scale=0.45]{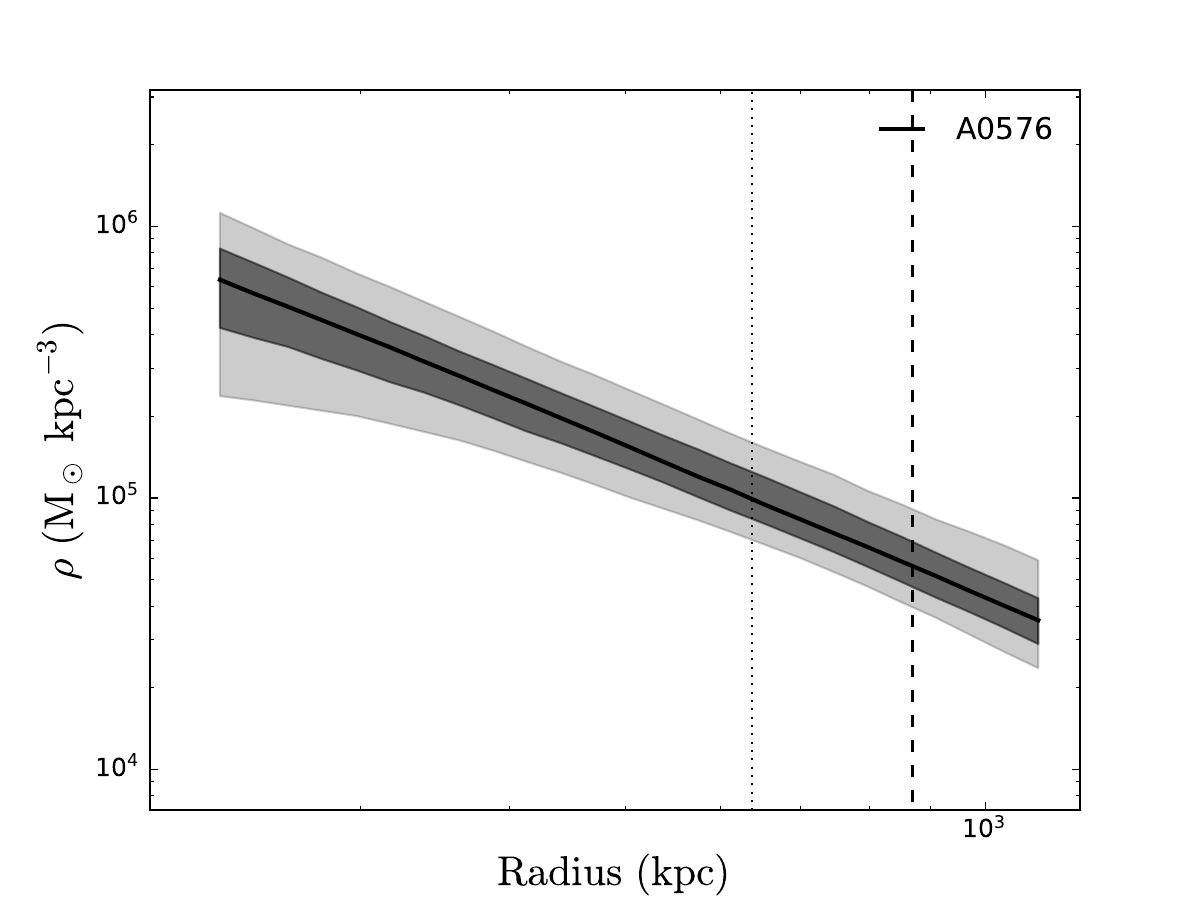}\\
    \includegraphics[scale=0.45]{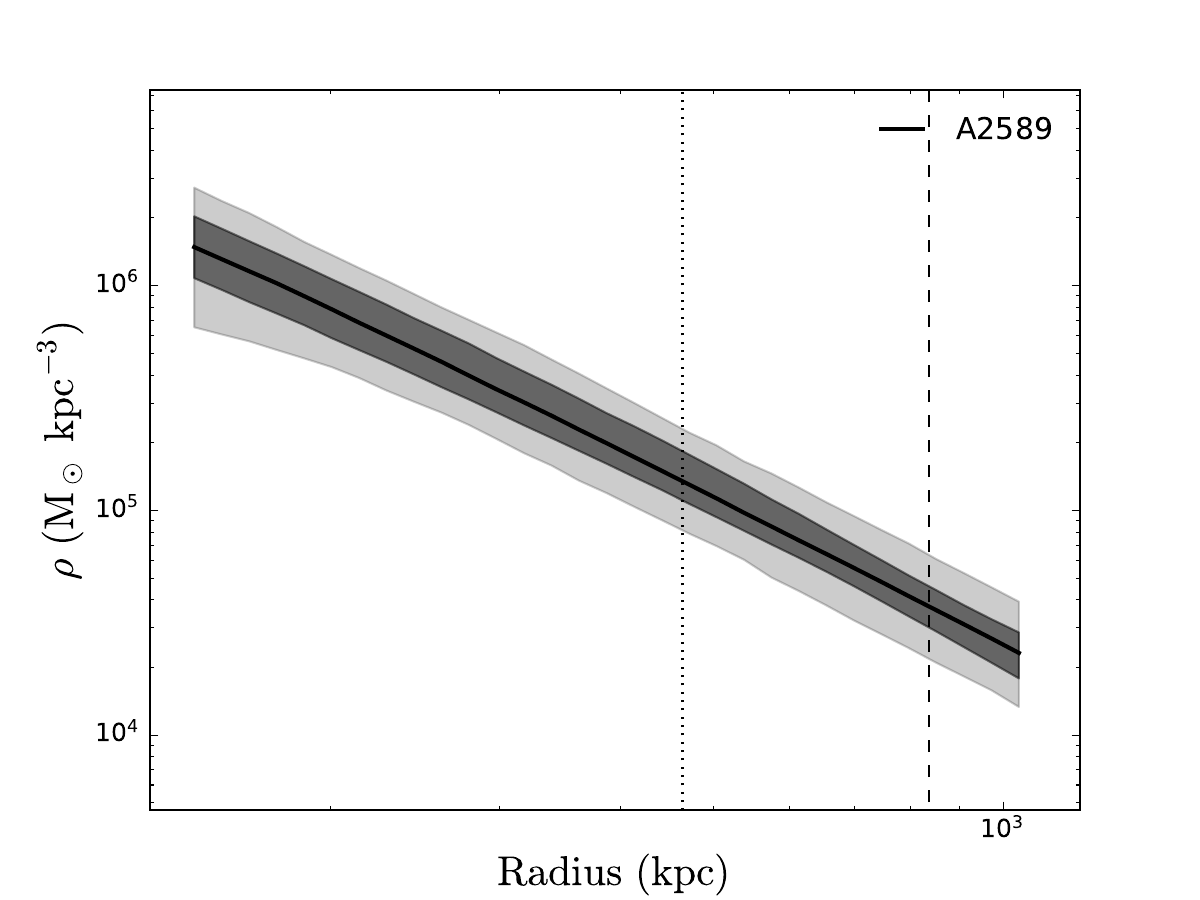}\includegraphics[scale=0.45]{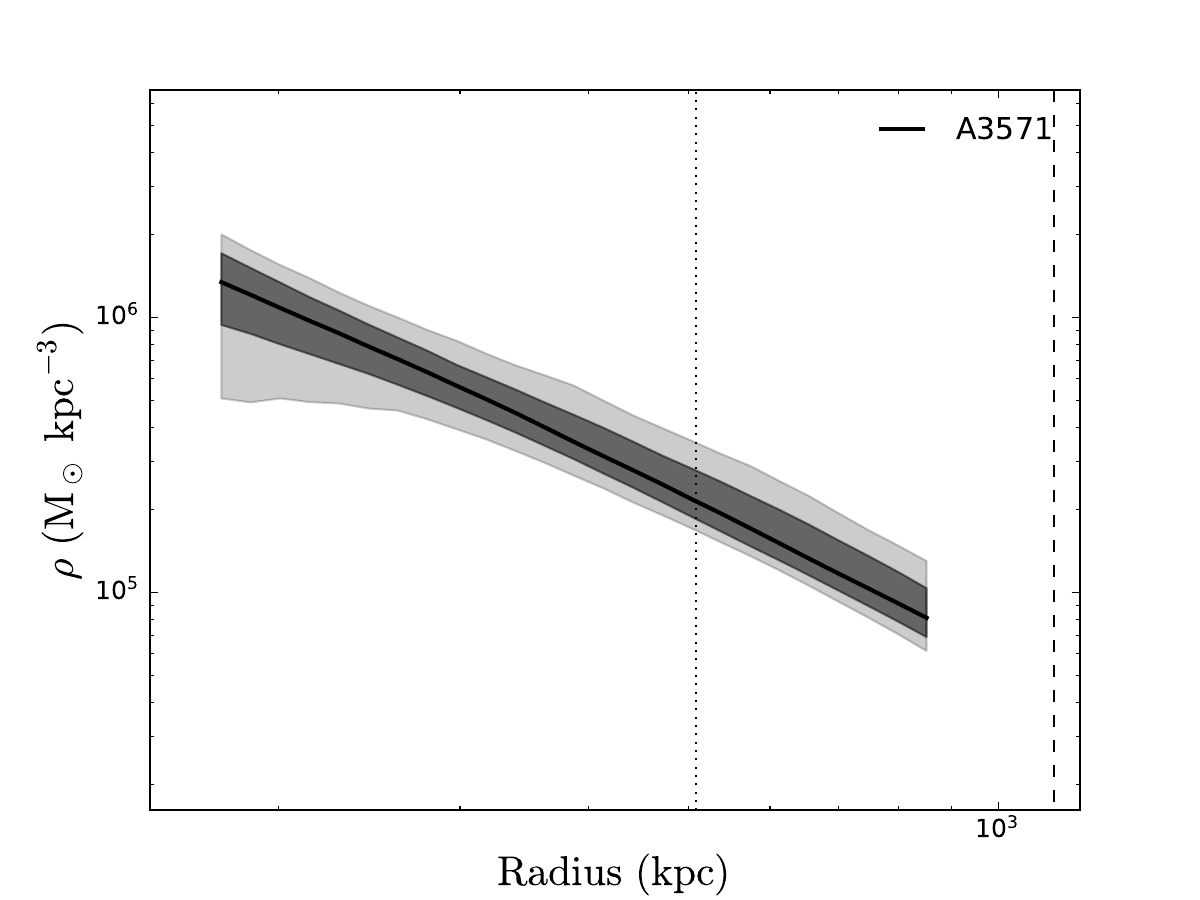}\\
    \caption{continued.}
    \label{fig:Densitycontinue}
\end{figure*}
\end{appendix}

\end{document}